\documentclass[a4paper,11pt]{article}
\pdfoutput=1 

\usepackage{jheppub} 
\usepackage{mathrsfs,graphicx,rotating,amsmath,amsfonts,mathtools,booktabs,amssymb,wasysym,caption}
\usepackage{hyperref}
\usepackage{slashed}
\usepackage{ulem}
\usepackage{natbib}
\usepackage[table,xcdraw,dvipsnames]{xcolor}
\usepackage{graphicx}
\usepackage{bbold}
\usepackage[utf8x]{inputenc}
\usepackage[english]{babel}
\usepackage{multirow,multicol}
\usepackage{epstopdf}
\usepackage{bbm}
\usepackage{appendix}
\usepackage{libertine}
\usepackage{wasysym}
\usepackage{empheq}
\usepackage{cancel}
\usepackage{enumitem}
\usepackage{mathrsfs}
\usepackage{lmodern}
\usepackage{tabularx}
\usepackage{multicol}
\usepackage{color}
\usepackage{mathtools}
\usepackage{xcolor}
\usepackage{bbold}
\usepackage{cancel}

\newcommand{\GD}{{\rm G}_{\rm D}}
\newcommand{\GW}{{\rm G}_{\rm W}}
\newcommand{\GSM}{{\rm G}_{\rm SM}}

\newcommand{\SUfiveGUT}{{\rm SU}(5)_{\rm GUT}}
\newcommand{\SUNDC}{{\rm SU}(N_{\rm DC})}

\newcommand{\SUm}{{\rm SU}(m)}
\newcommand{\UoneD}{{\rm U}(1)_{\rm D}}

\newcommand{\UonePQ}{{\rm U}(1)_{\rm PQ}}
\newcommand{\Uone}{{\rm U}(1)}
\newcommand{\SU}{\,{\rm SU}}
\newcommand{\SO}{\,{\rm SO}}
\newcommand{\U}{\,{\rm U}}
\newcommand{\gUV}{g_{UV}}
\newcommand{\deltaPQ}{\Delta_{\cancel{PQ}}}
\newcommand{\Nirr}{n_{irr}}

\makeatletter

\begin{document}

\title{Chiral models of composite axions and accidental Peccei-Quinn symmetry}
\author[a,b]{Roberto Contino,}
\author[c]{Alessandro Podo,}
\author[d]{Filippo Revello}

\affiliation[a]{Università di Roma La Sapienza,
Piazzale Aldo Moro 5, 00185 Roma, Italy}       
\affiliation[b]{Istituto Nazionale di Fisica Nucleare (INFN) - Sezione di Pisa 
\\ Polo Fibonacci Largo B. Pontecorvo, 3, 56127 Pisa, Italy}
\affiliation[c]{Department of Physics, Center for Theoretical Physics, Columbia University, 538 West 120th Street, New York, NY 10027, U.S.A.}
\affiliation[d]{Rudolf Peierls Centre for Theoretical Physics, Beecroft Building, Clarendon Laboratory,\\ Parks Road, University of Oxford, OX1 3PU, UK}
    
\abstract{%
We introduce a class of composite axion models that provide a natural solution to the strong CP problem, and possibly account for the observed dark matter abundance.
The QCD axion arises as a composite Nambu-Goldstone boson (NGB) from the dynamics of a chiral gauge theory with a strongly-interacting and confining \SU(N) factor and a weakly-interacting $\rm U(1)$, with no fundamental scalar fields.
The Peccei-Quinn (PQ) symmetry is accidental and all the mass scales are generated dynamically.
We analyze specific models where the PQ symmetry is broken only by operators of dimension 12 or higher. We also classify several other models where the PQ symmetry can be potentially protected up to the dimension 15 or 18 level.
Our framework can be easily extended to a scenario where the Standard Model~(SM) is unified into a simple gauge group, and we discuss the case of non-supersymmetric $\rm SU(5)$ unification.
The GUT models predict the existence of additional pseudo NGBs, parametrically lighter than the GUT and PQ scales, which could have an impact on the cosmological evolution and leave observable signatures.
We also clarify the selection rules under which higher-dimensional PQ-violating operators can generate a potential for the axion in the IR, and provide a discussion of the discrete symmetries in composite axion models associated to the number of domain walls. These results can be of general interest for composite axion models based on a QCD-like confining gauge group.}

\maketitle

\section{Introduction}
\label{sec:intro}

The strong CP problem can be elegantly solved by the QCD axion~\cite{Peccei:1977hh,Peccei:1977ur,Wilczek:1977pj,Weinberg:1977ma}, which assumes the existence of a spontaneously broken global symmetry - known as the Peccei-Quinn (PQ) symmetry - anomalous under $\SU(3)_{c}$. Crucially, QCD non-perturbative effects generate a potential for the axion such that the field at the minimum of its potential cancels the $\theta$ dependence exactly. In most of the axion models studied in the literature, the PQ symmetry is imposed by hand, leaving its UV origin unspecified.

In this work we introduce a class of models where the PQ symmetry arises as an accidental global invariance of a strongly-coupled chiral gauge theory, with no fundamental scalar fields in the PQ sector. The PQ symmetry is robustly protected by gauge invariance and the GUT dynamics does not induce PQ-breaking operators. The idea of an accidental PQ symmetry was first proposed in Ref.~\cite{Georgi:1981pu}, while that of a composite axion was put forward by Kim~\cite{Kim:1984pt} and then further studied in Refs.~\cite{Kaplan:1985dv,Choi:1985cb}. The setup we consider can be seen as an extension of Kim's model where however mass terms are \textit{forbidden} by a chiral U(1) gauge symmetry, so that the PQ invariance is accidental. The UV dark sector is similar to that of the Composite Dark Matter models introduced in \cite{Harigaya:2016rwr,Co:2016akw,Contino:2020god}; the crucial difference lies in the use of accidental symmetries to protect the PQ symmetry rather than species number. Moreover, the structural absence of mass terms automatically guarantees the vanishing of the CP-violating theta angle for the new strong dynamics.

The axion solution is notoriously fragile: additional UV contributions to the axion potential need to be extremely suppressed in order not to shift its minimum and spoil the cancellation of $\theta$.  
Simple arguments suggest that ultimately all global symmetries are explicitly broken by quantum gravity at the non-perturbative level. Planck suppressed operators with order one coefficients generally induce too large a PQ breaking, giving rise to the so-called axion quality problem~\cite{Kamionkowski:1992mf,Barr:1992qq,Holman:1992us,Ghigna:1992iv}. However, it is still unclear whether non-perturbative quantum gravity effects will be exponentially suppressed by a factor $e^{-c M_{Pl}/f_{a}}$, where $c$ is an $O(1)$ coefficient, as suggested by some semiclassical computations.
In the latter case, models with an accidental Peccei-Quinn symmetry can provide a fully satisfactory solution to the strong CP problem for axion decay constants $f_{a} \leq 1.2 \cdot 10^{16} \, \rm GeV$ -- see Ref.~\cite{Hebecker:2018ofv} for a review.
Some of the models analyzed in this work, in particular the minimal models of Section~\ref{sec:nf2}, naturally have an accidental PQ symmetry but do not solve the quality problem. They have a simple structure and could be relevant if the gravitational UV completion is weakly coupled and the dominant source of PQ breaking besides QCD comes from exponentially suppressed gravitational effects. In Sections~\ref{sec:nf2} and~\ref{sec:nf3} we present models with a slightly more elaborate structure that can also provide a natural structural solution to the axion quality problem. We analyze one particular such construction in Section~\ref{sec:nf3} in which the PQ symmetry is accidental up to dimension 12 operators.

Attempts to construct accidental composite axions solving the quality problem have been pioneered by Randall~\cite{Randall:1992ut}, who put forward a chiral model based on a new product gauge group $SU(n)\times SU(m)$,  where $SU(n)$ is assumed to get strong in the IR and confine, while $SU(m)$ remains weakly coupled and in the Higgs phase. The parameter space is significantly constrained by the requirement of having a confining $SU(n)$ dynamics and perturbative SM gauge couplings up to the Planck scale~\cite{Dobrescu:1996jp}.
Additional models based on non-abelian product gauge groups, which generalize Randall's construction and improve its quality, have been put forward in Refs.~\cite{Redi:2016esr,Vecchi:2021shj}. 
Supersymmetric composite axion models, also based on product gauge groups, have been instead constructed in Refs.~\cite{Lillard:2017cwx,Lillard:2018fdt}.
Other attempts to build composite axions and solve the quality problem by means of chiral gauge theories include the models of Ref.~\cite{Gavela:2018paw} and Ref.~\cite{Fukuda:2017ylt}. The first makes use of a chiral and confining simple gauge group, while the second exploits an axial $\Uone$ gauge group and is more similar to our approach. 
Holographic constructions dual to composite axion models have been analysed in Refs.~\cite{Choi:2003wr,Flacke:2006ad,Cox:2019rro,Bonnefoy:2020llz,Yamada:2021uze,Lee:2021slp}. 
Finally, several alternative approaches to the quality problem where the axion is a fundamental field have been proposed in the literature; for an incomplete list of works, see for example Refs.~\cite{Chun:1992bn,Cheng:2001ys,DiLuzio:2017tjx,Duerr:2017amf,Bonnefoy:2018ibr,Lee:2018yak,Hook:2019qoh,Ardu:2020qmo,Nakai:2021nyf,Darme:2021cxx,Bhattiprolu:2021rrj}.

In a string theoretic context, QCD axions can also arise through dimensional reduction of higher-form fields. Such axions are protected by an exact shift symmetry at the perturbative level, but the PQ symmetry is expected to be broken by non-perturbative effects such as non-QCD instantons. The size of these corrections depends on the value of the action of the relevant instanton configuration, and there are models in which they are small enough to have a viable QCD axion. This requirement, together with that of an axion decay constant parametrically smaller than the Planck scale, necessitates non generic UV features; nonetheless these models can provide fully consistent UV solutions to the axion quality problem~\cite{Svrcek:2006yi,Conlon:2006tq} (see also the recent work~\cite{Demirtas:2021gsq}). Our approach differs in that the solution to the quality problem is obtained as an IR property of the low-energy theory, the PQ symmetry being robustly protected by gauge invariance irrespectively of the details of the dynamics taking place at the Planck scale, provided that the theory admits a quantum gravity UV completion.

On a different note, the possible connection between the axion and an $\SU(5)$ Grand Unification dynamics has been first proposed in~\cite{Dine:1981rt,Wise:1981ry} and recently re-examined in Refs.~\cite{Boucenna:2017fna,DiLuzio:2018gqe,FileviezPerez:2019fku,FileviezPerez:2019ssf,Quevillon:2020aij}. All these models differ from the $\SU(5)$ model here described in two aspects: the Peccei-Quinn symmetry is not accidental and the dynamics responsible for its breaking is the GUT dynamics itself. We shall see instead that our setup makes it easy to realize the PQ symmetry as an accidental invariance of the GUT theory.
Along the same lines we follow, Refs.~\cite{Redi:2016esr,Vecchi:2021shj} considered $\SU(5)$ unified models with accidental composite axions. Additional attempts to build an accidental Peccei-Quinn invariance in theories of Grand Unification can be found in Refs.~\cite{Georgi:1981pu,DiLuzio:2020qio}. For an extensive review of QCD axion models we refer the reader to Ref.~\cite{DiLuzio:2020wdo}.

This article is organized as follows. The broad class of theories that will be considered in our analysis is introduced in Section~\ref{sec:mb}.
In Section \ref{sec:pq}, we study PQ violation from higher-dimensional operators in a general setting, and formulate a criterion to identify those operators which can affect the axion potential in the IR.
Models with $n_f=2$ and $n_f=3$ dark flavors are classified and analyzed in Sections~\ref{sec:nf2} and~\ref{sec:nf3} respectively, and their phenomenology is discussed in Section~\ref{sec:phenomenology}.
Appendix~\ref{appendix:discrete} provides a general analysis of discrete symmetries and the number of domain walls in composite axion models based on a vector-like confining gauge group. Appendix~\ref{appendix:QCD3} presents a simple QCD model with $n_f=3$ in detail, while Appendix~\ref{appendix:hq} contains a comprehensive list of charge assignments for $n_f=3$ yielding a high-quality axion. A comparison with Randall's model is also included in Appendix~\ref{appendix:Randall}. 

\section{Model building}\label{sec:mb}

We consider theories where the only UV degrees of freedom in the Peccei-Quinn sector are fermions and gauge fields\footnote{The SM sector can include scalars, such as the Higgs field or the ordinary heavy scalars of a grand unified UV completion.}. The gauge group is assumed to be of the form $\mathcal{G}=\GD \times \GSM$, where $\GSM$ is either the SM gauge group or one of its grand unified counterparts, and $\GD$ contains a confining subgroup that generates the Peccei-Quinn scale dynamically, giving rise to the composite axion. The PQ sector fulfills the following conditions:
\textit{i)}~it entails good theoretical control over the IR dynamics and its global symmetry breaking pattern;
\textit{ii)} it  leaves the SM gauge symmetries unbroken;
\textit{iii)} the representations of the new fermions under $\mathcal{G}$ are chiral.
In order to have control over the IR dynamics in the PQ sector, we assume that $\GD$ includes a confining dark color $\SUNDC$ subgroup under which the new fermions transform as fundamental plus anti-fundamental vectorlike representations. We take $N_{\rm DC} \geq 3$, so that the fundamental representation of $\SUNDC$ is complex. To ensure that the SM gauge symmetries are not broken spontaneously by $\SUNDC$ confinement we require that the new fermions transform also as vectorlike representations of $\GSM$. The chirality of the model then rests on the existence of an additional subgroup $\GW \subset \GD$ whose dynamics is perturbative at the $\SUNDC$ confinement scale. The case with $\GW = \SUm$ was first discussed in Ref.~\cite{Randall:1992ut} and then further analysed and generalised in Refs.~\cite{Dobrescu:1996jp,Redi:2016esr}. In this work we focus on the case in which the weak group is abelian, $\GW = \UoneD$. 

A general class of models fulfilling the conditions discussed above can be defined in terms of two sets of left-handed Weyl fermions transforming as
\begin{equation}
\label{eq:genmod}
\psi_i \sim (\square,p_i,r_i ), \qquad \chi_i \sim (\bar{\square},q_i, \bar{r}_i), \quad \quad i = 1,...,n_{f}
\end{equation}
under $\SUNDC \times \UoneD \times \GSM$. The $\UoneD$ charges $p_i$ and $q_i$ are defined such that $p_i \not = p_j$, $q_i \not = q_j$ for $i\not = j$ and $p_i \not = -q_i$. Therefore, $n_f$ is the number of different $\UoneD$ charges assigned to the $\psi$ (or equivalently to the $\chi$) fields, and the SM representations $r_{i}$ are in general reducible. The $\UoneD$ charges are chosen so that the representations are overall chiral. This requirement and the cancellation of $\UoneD$ gauge anomalies imply that the multiplicity of fermions $n_f$ has to be at least 2: $n_{f}\geq 2$. 

The most general renormalizable Lagrangian built from this field content comprises only kinetic terms, including a possible kinetic mixing between the hypercharge gauge boson and the $\UoneD$ gauge field:
\begin{equation}
\label{eq:BSMLagrangian}
\mathcal{L}_{\rm BSM} = -\dfrac{1}{4}  {\cal G}^{a}_{\mu\nu}{\cal G}^{a,\mu\nu} -\dfrac{1}{4}  F^D_{\mu\nu}F^{D\,\mu\nu} + \dfrac{\varepsilon}{2} F^D_{\mu\nu}B^{\mu\nu}+ \sum_{\kappa} \psi^{\dagger}_{\kappa} i D_{\mu} \bar\sigma^{\mu} \psi_{\kappa} + \sum_{\kappa} \chi^{\dagger}_{\kappa} i D_{\mu} \bar\sigma^{\mu} \chi_{\kappa}\, .
\end{equation}
Here ${\cal G}^a_{\mu\nu}$, $F^D_{\mu\nu}$ and $B_{\mu\nu}$ are the $\SUNDC$, $\UoneD$ and SM hypercharge field strengths respectively. No gauge invariant mass term or Yukawa coupling can be written due to the chiral structure of the model, and one can therefore set the $\SUNDC$ theta term to zero through an axial field redefinition. The label $\kappa$ runs over all irreducible representations of $\SUNDC \times \GSM \times \UoneD$, whose number we will denote by $\Nirr$.\footnote{One has $\Nirr = n_f$ if all the $r_i$ are irreducible representations.} The Lagrangian of Eq.~(\ref{eq:BSMLagrangian}) has a $[\Uone_V]^{\Nirr}\times [\Uone_A]^{\Nirr}$ accidental global symmetry at the classical level.\footnote{The accidental symmetry is bigger than that if any of the irreducible representations inside a given $r_i$ has a non-trivial multiplicity.} The $\Nirr$ vectorial $\Uone_V$ factors are linearly realized and each of them implies an accidentally conserved charge. One of the $\Nirr$ axial $\Uone_A$ factors is anomalous under $\SUNDC$, and one is gauged by $\UoneD$, whose vector boson acquires a mass through the Higgs phenomenon; the others are spontaneously broken by the $\SUNDC$ condensates $\langle\psi_{\kappa}\chi_{\kappa}\rangle$. The theory thus contains $(\Nirr-2)$ Nambu-Goldstone bosons (NGBs), one of which is identified with the axion. A necessary condition in order to have a composite axion is thus $\Nirr \geq 3$. The NGBs can acquire mass through higher-dimensional operators or from the QCD dynamics if they have anomalous couplings to the gluon field.

If the $\UoneD \times \GSM$ weak gauging is switched off, the global invariance of the Lagrangian of Eq.~(\ref{eq:BSMLagrangian})  gets enhanced to $\SU(N_{f})_L \times \SU(N_{f})_R \times U(1)_V$, where $N_{f}=\sum_{i=1}^{n_{f}} {\rm dim}(r_{i})$ is the multiplicity of Dirac fermions transforming in the fundamental representation of $\SUNDC$. Below the $\SUNDC$ confinement scale, the global symmetry group is spontaneously broken to the diagonal subgroup, and the low-energy theory is described by a set of $(N_f^2-1)$ NGBs. By turning on $\UoneD \times \GSM$, one of these NGBs becomes the longitudinal polarization of the massive $\UoneD$ gauge boson, and most of the other ones become pseudo and acquire a potential through perturbative loop effects. The only light degrees of freedom are the $(\Nirr-2)$ NGBs coming from the accidental symmetry discussed above.

This same model building scheme was thoroughly analyzed for $n_{f}=2$ in Ref.~\cite{Contino:2020god}, where the construction of Refs.~\cite{Harigaya:2016rwr,Co:2016akw} was generalized to new fermions charged under the SM gauge group. In the minimal setup of these works the choice of representations was such that $\Nirr=2$ and no light NGBs or axions were present in the infrared. The Dark Matter (DM) candidate was identified with the lightest pseudo NGB charged under an accidental species number, and the dynamical confinement scale was assumed to be of order $1-100\,$TeV, much smaller than a typical Peccei-Quinn scale. We shall discuss further the precise relation between this work and Ref.~\cite{Contino:2020god} in Section~\ref{sec:nf2}, where we classify and investigate models with $n_{f}=2$.

In the models studied in this work, the axion decay constant $f_a$ is determined in terms of the dark color confinement scale $\Lambda_{DC}$. The observation of the Supernova 1987A, together with the sharp prediction of the low-energy axion-nucleon couplings that follows in our models (see Eq.~(\ref{eq:lag_low}) and the discussion in Section~\ref{sec:phenomenology}), provides the most stringent lower bound on $f_a$, independently of its role as a cosmological relic: $f_{a} \gtrsim 4 \cdot 10^{8} \; \rm GeV$~\cite{Raffelt:2006cw} (see however Ref.~\cite{Bar:2019ifz}). It is well known that the axion can naturally reproduce the observed Dark Matter abundance through the misalignment mechanism for values of its decay constant $f_{a} \sim 10^{12}\,$GeV (for a recent numerical analysis see Refs.~\cite{Bonati:2015vqz,Borsanyi:2016ksw}). Motivated by these considerations, we will consider $f_{a} \sim 10^{9}\,$GeV  and $f_{a} \sim 10^{12}\,$GeV as benchmark values when analyzing our models.

\subsection{IR confinement and UV perturbativity}\label{sec:pt}

Our construction relies on the assumption that the $\SUNDC$ dynamics confines in the infrared. Since the new fermions have no mass term, $\SUNDC$ must be asymptotically free; at one loop this gives the bound $N_{\rm DC} \geq 2 N_{f}/ 11$. A stronger constraint $N_{f} < N_{\rm conf}$ is obtained by considering the lower boundary of the conformal window. The value of $N_{\rm conf}$ is a non-perturbative property that cannot be derived with perturbative computations. Lattice simulations suggest that $N_{\rm conf}\approx 12$ for an $\SU(3)$ gauge theory with fermions in the fundamental representation, see Ref.~\cite{DeGrand:2015zxa}.
In selecting our models we will use a naive linear extrapolation of this result to generic numbers of dark colors (see for instance Refs.~\cite{Appelquist:1996dq,Appelquist:1999hr,Poppitz:2009uq,Poppitz:2009tw} for some evidence in support of this assumption) and require
\begin{equation}
\label{eq:dc_coupling_conf}
N_f \leq 4 N_{\rm DC} \, .
\end{equation}
A second condition we impose on our theories is that the Standard Model gauge couplings should remain perturbative until the Planck scale. In practice, we require that there are no Landau poles below the Planck mass. Consider for example the QCD coupling; it is useful to define an effective number of new Dirac fermions in the fundamental of $\SU(3)_{\rm c}$ as
\begin{equation}
\Delta N_{f}^{(\rm QCD)} =2 N_{\rm DC} \sum_{i=1}^{n_{f}} \sum_{\alpha} T(r_{i}^{(\alpha)})\, ,
\end{equation}
where $r_{i}^{(\alpha)}$ are the irreducible representations of $\SU(3)_{\rm c}$ contained in $r_i$, and $T(r_{i}^{(\alpha)})$ are their corresponding Dynkin indices.\footnote{We normalize the Dynkin index so that it is equal to $1/2$ for the fundamental representation.} Running the QCD coupling up to the Planck scale, the perturbativity requirement implies the following condition:
\begin{equation}
\label{eq:qcd_coupling}
\Delta N_{f}^{(\rm QCD)} \leq \dfrac{3}{2} \left(\log\dfrac{M_{\rm Pl}}{\Lambda_{\rm DC}} \right)^{-1} \left(\dfrac{2\pi}{\alpha_{s}(m_{Z})}+ b^{(3)}_{\rm SM} \log\dfrac{M_{\rm Pl}}{m_{Z}}  \right),
\end{equation}
where $\alpha_{s}(m_{Z})=0.118$ is the QCD coupling constant at the Z pole~\cite{Zyla:2020zbs}, $m_{Z}= 91\, \rm GeV$ is the $Z$ mass and $b^{(3)}_{\rm SM}=7$ is the SM one-loop beta function coefficient.\footnote{More precisely, one should integrate out the top quark at its mass scale $m_t$ and use $b^{(3)\,\prime}_{\rm SM}=7+2/3$ to run from $m_Z$ to $m_t$. This leads to a small correction to Eq.~(\ref{eq:qcd_coupling}) that we neglect for simplicity.}
Numerically, this gives $\Delta N_{f}^{(\rm QCD)} \lesssim 26.7$ for $\Lambda_{\rm DC} = 10^{11}\, \rm GeV$ and
$\Delta N_{f}^{(\rm QCD)} \lesssim 35.6$ for $\Lambda_{\rm DC} = 10^{13}\, \rm GeV$. More in general, since $\Delta N_{f}^{(\rm QCD)}  \propto N_{\rm DC}$ and $\Lambda_{\rm DC} \propto \sqrt{N_{\rm DC}} f_a$ (as it will be shown in the following), the perturbativity condition of Eq.~(\ref{eq:qcd_coupling})  sets a lower bound on $f_a$ for fixed $N_{\rm DC}$, or equivalently an upper bound on $N_{\rm DC}$ for a given $f_a$.
Perturbativity of the hypercharge coupling up to the Planck scale similarly requires a bound
\begin{equation}
\label{eq:Y_coupling}
2 N_{\rm DC} \sum_i \sum_{\alpha} y_{i,\alpha}^2 \text{dim}(r_i^{(\alpha)}) \leq \frac{3}{4} \left(\log\dfrac{M_{\rm Pl}}{\Lambda_{\rm DC}} \right)^{-1} \left(\dfrac{2\pi}{\alpha_{s}(m_{Z})}+ b^{(1)}_{\rm SM} \log\dfrac{M_{\rm Pl}}{m_{Z}}  \right),
\end{equation}
where $y_{i,\alpha}$ is the hypercharge of the representation $r_i^{(\alpha)}$ and $b^{(1)}_{\rm SM} = -41/6$.

In the following, we will focus on models satisfying the conditions of Eqs.~(\ref{eq:dc_coupling_conf}),~(\ref{eq:qcd_coupling}) and~(\ref{eq:Y_coupling}).

\section{Peccei-Quinn symmetry and axion quality}
\label{sec:pq}

As highlighted in Section~\ref{sec:intro}, the existence of higher-dimensional PQ-violating operators poses a significant challenge to the axion solution of the strong CP problem. A simple estimate gives an idea of how serious this problem can be. If the PQ breaking effect comes from a single insertion of a local operator generated at the Planck scale, one expects a shift of the value of $\theta$ at the minimum of the potential of order
\begin{equation}
\label{eq:pqe}
\Delta\theta \approx |c| \, \phi_{CP} \left(\frac{M_{\rm Pl}}{\Lambda_{\rm QCD}}\right)^4 \left( \frac{f_a}{M_{\rm Pl}} \right)^{\deltaPQ},
\end{equation}
where
$\Delta_{\cancel{PQ}}$ is the dimension of the operator, $c$ is a dimensionless effective coefficient and $\phi_{CP}$ is a $CP$-violating phase. The current experimental bound on the electric dipole moment (EDM) of the neutron~\cite{nEDM:2020crw} requires $\Delta\theta \lesssim 10^{-10}$, which translates into a lower bound on the dimension of the PQ breaking operator
\begin{equation}\label{eq:dpq}
\deltaPQ >  \dfrac{10+\log_{10}\!\left(|c| \, \phi_{CP}  \dfrac{M_{\rm Pl}^{4}}{\Lambda_{\rm QCD}^{4}}\right)}{\log_{10}\!\left(\dfrac{M_{\rm Pl}}{f_{a}}\right)}\, .
\end{equation}
Assuming $|c|, \phi_{CP}\sim 1$ and using $M_{\rm Pl} \sim 10^{19}\,$GeV, $\Lambda_{\rm QCD} \sim 100\,$MeV, this gives $\deltaPQ \gtrsim 9$ for $f_{a} \sim 10^{9}\,$GeV. Values of the PQ scale of order $f_{a} \sim 10^{12} \,{\rm GeV}$, which allow the axion to reproduce the observed dark matter abundance naturally through the misalignment mechanism, lead to a stronger bound $\deltaPQ \gtrsim 13$. In theories of composite axion, for a purely fermionic PQ-violating operator generated through diagrams with $\ell$~loops, one expects the effective coefficient to scale as $c\sim \gUV^{2} (4\pi/\gUV)^{\deltaPQ/3-2\ell}$, where $\gUV$ is a UV coupling. This corresponds to values of $c$ larger than 1, hence to a stronger constraint on $\deltaPQ$, unless $\ell > \deltaPQ/6-1$ and $\gUV$ is sufficiently small.\footnote{The more accurate estimate of $\Delta\theta$ in composite axion theories performed by Ref.~\cite{Vecchi:2021shj} suggests that the actual constraint might be even stronger due to the presence of very large multiplicity factors. These are smaller in our models compared to those of Ref.~\cite{Vecchi:2021shj} though still quite large. For $\deltaPQ =12$, including such factors leads to an upper bound on $f_a$ stronger by roughly one order of magnitude.} Such strong bounds on the dimension $\deltaPQ$ make it clear that obtaining a natural resolution of the axion quality problem is not an easy task.

The problem is even more serious if PQ-violating operators are generated at energies lower than the Planck scale by perturbative dynamics. In this case the minimum values of $\Delta_{\cancel{PQ}}$ required to preserve the axion solution are larger. Theories of Grand Unification where the PQ symmetry does not commute with the GUT group are an example of this situation. For GUT scales of order $10^{17}\,$GeV one in general needs $\Delta_{\cancel{PQ}} \gtrsim 10$ for $f_{a} \sim 10^{9}\,$GeV, and $\Delta_{\cancel{PQ}} \gtrsim 16$ for $f_{a} \sim 10^{12}\,$GeV. Compatibility of the axion solution with Grand Unification thus imposes non trivial constraints on the structure of the theory.

In our analysis of candidate models with accidental PQ symmetry below the Planck scale, we will take the minimum dimensionalities $\Delta_{\cancel{PQ}} = 9$ and $\Delta_{\cancel{PQ}} = 12$ as targets for a solution of the axion quality problem for $f_a = 10^9\,$GeV and $f_a = 10^{12}\,$GeV respectively.
We choose these values considering the high sensitivity of Eq.~(\ref{eq:dpq}) on the value of the PQ scale and the possibility of a (loop) suppression of the effective coefficient $c$ and of the $CP$-violating phase $\phi_{CP}$. 

Throughout this work we will compute the dimensionality of operators by assuming that the theory is weakly coupled at the UV scale and that the dark color force gets strong near the confinement scale $\Lambda_{\rm DC}$. In this case, the scaling dimension of operators is equal to the classical dimension up to small corrections. On the other hand, if the dark color dynamics has a walking (as opposed to running) behavior, i.e. it remains strongly coupled all the way up to the UV cutoff scale or slightly below, then the RG evolution above $\Lambda_{\rm DC}$ is characterized by scaling dimensions that might be significantly larger than the classical ones, thus ameliorating or even erasing the axion quality problem. A dynamics of this kind can for example arise if $N_{\rm DC}$ is such that dark color confines but is close to the conformal window. In this case the RG evolution between the UV scale, where the PQ breaking effects are generated, and the infrared confined phase below $\Lambda_{\rm DC}$ is characterized by a strongly-coupled flow lingering close to a pair of fixed points at complex coupling~\cite{Kaplan:2009kr,Gorbenko:2018ncu}. While the scaling dimension of PQ-violating operators cannot be computed analytically at strong coupling, and the predictivity of our theoretical construction is lost in this limit,\footnote{One could however use supersymmetry to gain analytic control over the strong dynamics, see for example Ref.~\cite{Nakai:2021nyf} for a model of this kind.} a scenario of this kind could be relevant for the resolution of the quality problem and is worth being investigated. Interestingly, realizations of this idea in the context of warped extra dimensions could provide calculability and lead to a concrete candidate model, see for example Ref.~\cite{Cox:2019rro}.

\subsection{General remarks on PQ violation: a spurion analysis}
\label{sec:spurion}

As already noticed by Ref.~\cite{Dobrescu:1996jp}, not all PQ-violating operators can generate a potential for the axion at low energy. Since identifying the dangerous operators is obviously crucial to single out viable theories, in this section we perform a detailed analysis based on selection rules and spurion arguments. Our results apply to any $\SUNDC \times \GW \times \GSM$ gauge theory with Weyl fermions $\psi$ and $\chi$ transforming as vectorlike representations of $\SUNDC \times \GSM$.

A generic PQ-violating operator ${\cal O}_{\cancel{PQ}}$ induces a potential for the axion if the matrix element $\langle \psi_{a,\varphi}| {\cal O}_{\cancel{PQ}} | 0\rangle$ is non vanishing. Here $|\psi_{a,\varphi}\rangle$ is an arbitrary state containing axions and possibly other SM-neutral NGBs, denoted by $\varphi$, with zero momentum.\footnote{While we are ultimately interested in a potential for just the axion field, we conservatively consider the possibility of generating a mixed potential with other SM-neutral NGBs, which can in turn lead to an axion potential at energies smaller than the mass of the $\varphi$'s.} Therefore, ${\cal O}_{\cancel{PQ}}$ is an interpolating operator for the state $|\psi_{a,\varphi}\rangle$ and must have its same quantum numbers. 

In order to determine the form of ${\cal O}_{\cancel{PQ}}$ implied by this requirement, it is useful to first neglect the effects of the weak gauging of $\GW\times \GSM$. In this limit, the global symmetry is $G_F = \SU(N_{f})_L \times \SU(N_{f})_R \times \U(1)_V$, under which the fermions transform as
\begin{equation}
\label{eq:trasf}
\psi^{i_L} \rightarrow e^{i \alpha}\, L^{i_L}_{j_L} \, \psi^{j_L},\quad \quad \quad  \chi_{i_R} \rightarrow e^{-i\alpha} \, (R^*)_{i_R}^{j_R} \, \chi_{j_R} \, .
\end{equation}
Any $G_F$-violating operator made of the $\psi$ and $\chi$ fields can thus be characterized by a spurionic tensorial structure
\begin{equation}
T^{\{i_L\} \{j_R\} }_{\, \{k_L\} \{ l_R\} }  \, ,
\end{equation}
where $\{ n \}$ denotes a set of $n$ indices. At energies below the $\SUNDC$ confinement scale, $G_F$ is spontaneously broken to $\SU(N_{f})_V \times \U(1)_V$, and the corresponding NGBs are conveniently described by the field $\Sigma = \exp(i \pi^{\hat a} T^{\hat a})$. The latter transforms as $\Sigma \to L \Sigma R^\dagger$ and thus carries one upper index of $\SU(N_{f})_L$ plus one lower index of $\SU(N_{f})_R$. A state of zero-momentum NGBs can be excited by the chiral field $\Sigma$ and, more in general, by effective operators in the low-energy theory that are polynomials of $\Sigma$ and $\Sigma^\dagger$. Since we are interested in states with axions and light NGBs, we need to consider exponentials of their fields. Because $\Sigma$ is unitary, each one of its indices will contribute to determine the spurionic structure of the interpolating effective operator (no $G_F$ singlet contractions are possible). The latter must therefore have an equal number of upper left and lower right indices, as well as an equal number of lower left and upper right indices. In other words, the spurionic structure of an interpolating operator in the low-energy theory must be of the form $T^{\{i\} \{j\} }_{\, \{j\} \{ i\} }$; clearly, the interpolating operator in the UV theory must have the same spurionic structure. Furthermore, any such operator must have zero $\U(1)_V$ charge, i.e. be a `mesonic' operator. This implies that ${\cal O}_{\cancel{PQ}}$ is necessarily a polynomial of the $\SUNDC$-singlet bilinears $(\psi_r \chi_{\bar r})$ and $(\psi_r \chi_{\bar r})^\dagger$ (whose flavor indices are left understood), where $\psi_r$ and $\chi_r$ denote fields transforming as the irreducible SM representation $r$. Such fermionic bilinears are exactly those acquiring a vacuum expectation value at the $\SUNDC$ confinement scale. The same conclusion was obtained by Dobrescu in Ref.~\cite{Dobrescu:1996jp}.

Let us now consider the effects of the weak gauging of $\GW\times \GSM$. The latter breaks explicitly the flavor group $G_F$ down to $[\Uone_V]^{\Nirr}\times [\Uone_A]^{\Nirr}$, as we have already discussed, where the axial $[\Uone_A]^{\Nirr}$ is non-linearly realized.
The state $|\psi_{a,\varphi}\rangle$ is neutral under all the vectorial $\Uone$'s and under the SM group, while it transforms non-trivially under the PQ symmetry. Since the same quantum numbers must also characterize ${\cal O}_{\cancel{PQ}}$, we conclude that

\vspace{0.2cm}
\textit{A gauge-invariant PQ-violating operator ${\cal O}_{\cancel{PQ}}$ can generate a potential for the axion at low energy only if it has vanishing vectorial charges.}
\vspace{0.2cm}

Operators satisfying this requirement are polynomials of the bilinears $(\psi_r \chi_{\bar r})$, $(\psi_r \chi_{\bar r})^\dagger$, $(\psi^\dagger_r \psi_{r})$ and $(\chi^\dagger_{\bar r} \chi_{\bar r})$. Hence, operators built of $(\psi^\dagger_r \psi_{r})$ and $(\chi^\dagger_{\bar r} \chi_{\bar r})$ factors are also allowed as a consequence of the weak gauging.\footnote{Notice that $(\psi^\dagger_r \psi_{r})$ and $(\chi^\dagger_{\bar r} \chi_{\bar r})$ transform as vectors under the Lorentz group. If in a given local operator they appear with gauge indices contracted to form a singlet of $G_{\rm SM}$, then a new operator can be constructed with lower dimensionality where each bilinear is replaced by a covariant derivative.} The axion potential in that case will be suppressed by powers of the weak gauge couplings. 

It is instructive to analyze some examples in the more familiar case of QCD, where a similar selection rule characterizes the operators that can generate a potential for the pions. Consider for example the local operator $O =(u d^c)(s u^c)$ in three-flavor QCD where electromagnetism plays the role of the weak gauging.\footnote{Here $q^c$ denotes the complex conjugate of the right-handed quark $q$. Therefore, $q^c$ is a left-handed field transforming as an anti-fundamental under color.} In the limit of vanishing up and down quark masses, $\pi^0$ is an exact NGB at the renormalizable level, and it is interesting to ask whether it gets a potential and a mass from insertions of $O$. The latter explicitly breaks the  $U(1)_{3A}$ axial symmetry, but has non zero isospin (i.e. charge under the vectorial $U(1)_{3V}$) and thus cannot generate any potential for $\pi^0$.  More in detail, one can see that in the low-energy theory $O$ gives rise to non-derivative interactions among $\pi^0$ and the other (pseudo) NGBs; these interactions violate isospin by $+1/2$ units. Since the rest of the dynamics conserves isospin, there are no diagrams at any loop order in chiral perturbation theory that generate a potential for $\pi^0$. Another instructive example is that of the operator $O = (u u^c)(u^\dagger \bar\sigma^\mu u)^2$ in two-flavor QCD with vanishing quark masses plus electromagnetism. By conveniently parametrizing $\Sigma = \exp[i(\pi^+ \sigma^+ + \pi^- \sigma^-)] \exp(i \pi^0 \sigma^3)$, it is simple to see that at low energy $O \sim \cos\pi^0 \cos\sqrt{\pi^+ \pi^-} \big( \partial_\mu \pi^0 + i \pi^+ \!\overleftrightarrow{\partial}_{\!\!\mu} \pi^-/2 + \dots \big)^2$. A potential for $\pi^0$ is generated at leading order by diagrams with two loops of charged pions. In the limit in which electromagnetism is switched off, the charged pion becomes massless and these loops vanish identically (one can for example use dimensional regularization to preserve chiral invariance). Therefore, the operator $O$ generates a potential only in presence of the electromagnetic (weak) gauging; this is expected since it contains factors of $(u^\dagger u)$.

So far we have implicitly assumed that the axion potential is generated by single insertions of a local PQ-violating operator. However, multiple insertions of local operators are also relevant and their importance was not emphasized in the previous literature. On dimensional grounds, the insertion of $N$ local operators of dimensions $d_i$ gives a contribution equivalent to the insertion of a single operator of effective dimension
\begin{equation} \label{eq:mult_ins}
d_{\text{eff}} = \sum_{i=1}^N d_i - 4(N-1)\, .
\end{equation}
This is smaller than the dimension of the operator built out of their product by $4(N-1)$. It is then possible for multiple insertions to give the dominant contribution to the axion potential. Indeed, while the product of local operators will have to satisfy the selection rule discussed above and thus be a polynomial of  $(\psi_r \chi_{\bar r})$, $(\psi_r \chi_{\bar r})^\dagger$, $(\psi^\dagger_r \psi_{r})$ and $(\chi^\dagger_{\bar r} \chi_{\bar r})$, the individual operators need only to violate flavor (otherwise they play no role) and at least one of them needs to break the PQ symmetry. Notice that $\GSM$ invariance of each individual operator can be achieved by adding SM or GUT fields. In sec.~\ref{ssc:exso} we will discuss one example of a theory where the existence of dimension-6 operators prevents the resolution of the quality problem despite their single insertion cannot generate an axion potential.

From these considerations it should be clear that a complete classification of the higher-dimensional operators capable of spoiling the axion quality is not an easy task, even for a specific model. One would like to determine, for each given model, the smallest dimension $\Delta _{\cancel{PQ}}$ at which an axion potential is generated. This can be defined as the minimum over
\begin{itemize}
\item[(1)] the dimensions of all gauge-invariant, PQ-violating local operators that are polynomials of  $(\psi_r \chi_{\bar r})$, $(\psi_r \chi_{\bar r})^\dagger$, $(\psi^\dagger_r \psi_{r})$, $(\chi^\dagger_{\bar r} \chi_{\bar r})$; and
\item[(2)] the effective dimensions of all gauge-invariant, PQ-violating polynomials of $(\psi_r \chi_{\bar r})$, $(\psi_r \chi_{\bar r})^\dagger$, $(\psi^\dagger_r \psi_{r})$, $(\chi^\dagger_{\bar r} \chi_{\bar r})$ that can be formed as the product of gauge-invariant, flavor-violating local operators.
\end{itemize}
Non-local operators of type (2) arise from multiple insertions of local operators. We will thus consider a model able to resolve the axion quality problem for some given physically-allowed value of $f_a$ if the smallest PQ-violating dimension $\Delta _{\cancel{PQ}}$ defined as above satisfies the bound of Eq.~(\ref{eq:dpq}). In particular, we will require $\Delta _{\cancel{PQ}} \geq 9$ for $f_a\sim 10^9\,$GeV and $\Delta _{\cancel{PQ}} \geq 12$ for $f_a\sim 10^{12}\,$GeV.

Notice that multiple insertions of operators with dimension bigger than $(d+4)/2$ are always more suppressed than a single insertion of a dimension $d$ operator. This means that non-local operators of type (2) are not relevant to determine $\Delta _{\cancel{PQ}}$ if they have dimension larger than $(\Delta _{\cancel{PQ}}+4)/2$. The calculation of $\Delta _{\cancel{PQ}}$ is therefore simplified if multiple insertions can be neglected based on the minimum dimensionality of flavor-violating operators. The number of $n_f=2$ theories of this kind is for example reported in Tables~\ref{tab:hqax} and~\ref{tab:hqax3}.

\section{Models with $n_{f}=2$} 
\label{sec:nf2}

Models with $n_{f}=2$ and $r_{i}$ irreducible have been studied and classified in Ref.~\cite{Contino:2020god}. With reference to the general structure of Table~\ref{tab:nf2}, the most general charge assignments that are vector-like under $\SUNDC \times \GSM$ and overall chiral in the case of irreducible $r_{i}$ are as follows: $p_{1}=-p_{2}=1$, $q_{2}=-q_{1}=q$, with $r_{1}=r_{2} \equiv r$ and $q \in [0,1)$ (Type I); $p_{1}=-p_{2}=1$, $q_{2}=-q_{1}=q$, with $r_{1}=\bar{r}_{2} \equiv r$ and $q \in [-1,1)$ (Type II); $p_{1}=-q_{2}=1$, $p_{2}=-q_{1}=q$, with $r_{1}=\bar{r}_{2} \equiv r$ and $q \in [0,1)$ (Type III). The only phenomenologically viable models consistent with values of the confinement scale $\Lambda_{\rm DC} \sim 1-100 \, \rm TeV$ were identified in Ref.~\cite{Contino:2020god}  to be those with $\SU(3)_c\times\SU(2)_{EW}\times \U(1)_Y$ representations $r= (\mathbf{1},\mathbf{2})_{0}$ and $r= (\mathbf{3},\mathbf{1})_{0}$.
\begin{table}[t]
\centering
\begin{tabular}{lccc}
& $\SUNDC$ & $\UoneD$ & $\GSM$  \\
\cline{1-4} 
\rule{0pt}{2.4ex}$\psi_1$ & $\square$ &  $p_{1}$ & $r_1$ \\
$\psi_2$ & $\square$ &  $p_{2}$ & $r_2$\\         
\cline{1-4}        
\rule{0pt}{2.4ex}$\chi_1$ & $\bar{\square}$ &  $q_{1}$  & $\bar{r}_1$ \\        
$\chi_2$ & $\bar{\square}$ &  $q_{2}$  & $\bar{r}_2$\\ 
\end{tabular}
\caption{\it General structure of accidental axion models with $n_{f}=2$. The possible charge assignments $\{p_{1},p_{2},q_{1},q_{2}\}$ are restricted by the cancellation of gauge anomalies as discussed in the main text. The group $\GSM$ is either the SM gauge group or one of its GUT extensions, and the possible representations $r_{i}$ are classified in Section~\ref{sec:nf2_class}.}
\label{tab:nf2}
\end{table}
Theories with reducible $r_i$ were found to be strongly constrained by the requirement of perturbativity. In practice, the only non-trivial candidate was a Type-II theory with $r$ equal to the direct sum of one electroweak doublet plus one color triplet; by combining these two irreps into a fundamental representation of a unified $\SU(5)$ group one could obtain a GUT model. However, such a theory turns out to be not phenomenologically viable in the minimal scenario considered in Ref.~\cite{Contino:2020god}. Indeed, it contains two light pseudo NGBs, one of which is an axion-like particle. The corresponding PQ symmetry does not commute with $\SU(5)$ and is explicitly broken by dimension-6 four-fermion operators generated at the GUT scale. Such effects induce an extra contribution to the potential of the axion-like NGB that dominates over the QCD one for values of the strong dynamical scale above $\sim 10^8\,$GeV, thus spoiling the resolution of the strong CP problem. Lower values of the strong dynamical scale are instead excluded by astrophysical bounds on axion-like particles.

When raising the dynamical scale to values above $\sim 10^8\,$GeV, however, new choices of reducible representations $r_i$ become allowed by perturbativity, and it is possible to construct realistic axion models. This requires having $\Nirr\geq 3$, as discussed in Section~\ref{sec:mb}, and one way to obtain this is to consider $n_{f}=2$ with $r_{i}$ reducible. The simplest possibility is to start from the minimal model of Ref.~\cite{Contino:2020god} with QCD triplets and enlarge each representation by the addition of one SM singlet. We follow this route and present this minimal QCD model first and then its unified version. These models turn out to have $\Delta_{\cancel{PQ}} = 6$ and therefore do not solve the quality problem, but it is nonetheless very instructive to analyze them. We then generalize this construction in Section~\ref{sec:nf2_class}, where we classify all possible models with $n_{f}=2$ and generic reducible representations. Our analysis shows that these additional models have a chance of solving the quality problem.

\subsection{A Minimal QCD model}\label{sec:qcdm}

We consider a Type-II model with the charge assignment detailed in Table~\ref{tab:model} and $r=(\mathbf{3},\mathbf{1})_{+y}$ under $\GSM=\SU(3)_c\times\SU(2)_{EW}\times \U(1)_Y$.
\begin{table}
\centering
\begin{tabular}{lccc}
& $\SUNDC$ & $\UoneD$ & $\rm \GSM$  \\
\cline{1-4} 
\rule{0pt}{2.4ex}$\psi_1$ & $\square$ &  $+1$ & $1$ \\
$\psi_2$ & $\square$ &  $-1$ & $1$ \\         
$\psi_3$ & $\square$ &  $+1$ & $r$ \\
$\psi_4$ & $\square$ &  $-1$ & $\bar{r}$ \\  
\cline{1-4}        
\rule{0pt}{2.4ex}$\chi_1$ & $\bar{\square}$ &  $-q$  & $1$ \\        
$\chi_2$ & $\bar{\square}$ &  $+q$  & $1$ \\ 
$\chi_3$ & $\bar{\square}$ &  $-q$   & $\bar{r}$ \\
$\chi_4$ & $\bar{\square}$ &  $+q$  & $r$ \\
\end{tabular}
\caption{\it Minimal accidental axion models with $n_f=2$. All the fields $\psi_{i}$, $\chi_{i}$ are left-handed Weyl fermions, and the parameter $q$ is a rational number in the interval $(-1,1)$. The QCD and GUT models discussed in the text correspond to $\GSM=\SU(3)_c\times\SU(2)_{EW}\times \U(1)_Y$ with $r=(\mathbf{3},\mathbf{1})_{+y}$, and to $\GSM=\SU(5)_{\rm GUT}$ with $r=\mathbf{5}$, respectively.}
\label{tab:model}
\end{table}
For this choice, the SM triplets can be assigned a non-zero hypercharge $y$, which modifies the low-energy axion-photon coupling as we shall see. Type I and Type III models can be also constructed for $y=0$ and lead to the same low-energy phenomenology.\footnote{For non-vanishing hypercharge, Type I and Type III models have mixed gauge anomalies, see Ref.~\cite{Contino:2020god}.} They differ from Type-II models only for the spectrum of heavy pseudo NGBs and other resonances with mass of order $\Lambda_{\rm DC}$, which appear to be inaccessible with current and near-future experiments. For any rational value of the parameter $q$ in the interval $(-1,1)$, the dark fermions have chiral representations and no gauge invariant mass term or Yukawa coupling can be written.\footnote{For $q=-1$ the fields singlet under $\GSM$ form vectorlike representations. Therefore, differently from the case with $n_f = \Nirr=2$ studied in Ref.~\cite{Contino:2020god}, this value of $q$ must be discarded.} 

At the classical level the model has a ${\rm U}(1)_L^{4} \times {\rm U}(1)_R^{4}$ global symmetry, corresponding to the group of phase transformations of each of the fields $\psi_{i}$ and $\chi_i^c$ ($\chi^c$ is the conjugate of $\chi$ and a right-handed Weyl fermion). At confinement the pattern of symmetry breaking is 
\begin{equation}\label{eq:sb1}
{\rm U}(1)_L^{4} \times {\rm U}(1)_R^{4} \longrightarrow {\rm U}(1)_V^{4}\, ,
\end{equation}
giving four spontaneously broken ${\rm U}(1)$ axial factors. One of them is anomalous under $\SUNDC$ and one is gauged by $\UoneD$. We are thus left with two broken generators, associated with two Nambu-Goldstone bosons. These are neutral under the unbroken vectorial ${\rm U}(1)_V$ symmetries, and will be denoted by $s$ and $a$ in the following.

Defining $\Psi_{L} = (\psi_{1} \, \psi_{2} \, \psi_{3} \, \psi_{4})$ and $\Psi_{R} =(\chi^{c}_{1} \; \chi^{c}_{2} \; \chi^{c}_{3} \; \chi^{c}_{4} )$,  $s$ and $a$ are interpolated by the axial currents
\begin{equation}
J_{s}^{\mu} = \overline{\Psi} \gamma^{\mu} \gamma^{5} Q_{s} \Psi\, , \qquad
J_{a}^{\mu} = \overline{\Psi} \gamma^{\mu} \gamma^{5} Q_{a} \Psi\, , 
\end{equation}
with
\begin{equation}\label{eq:gen}
Q_{s} = {\rm diag}(3,-3,-\mathbb{1},\mathbb{1})\, , \qquad
Q_{a} = {\rm diag}(-3,-3,\mathbb{1},\mathbb{1}) \, .
\end{equation}
The current $J^{\mu}_{a}$ has an anomaly in the background of SM gauge fields:
\begin{equation}
\langle \partial_{\mu} J_{a}^{\mu} \rangle = - \dfrac{\alpha_{3}}{4\pi} A_{3}\, G \tilde{G}- \dfrac{\alpha_{Y}}{4\pi} A_{Y} \, B\tilde{B},
\end{equation}
where $G_{\mu \nu}$ and $B_{\mu \nu}$ are the QCD gluon and hypercharge field strengths respectively, the subscripts $3$ and $Y$ refer to the QCD and hypercharge gauge couplings, and dual field strengths are defined as $\tilde{F}^{\mu\nu} = \epsilon^{\mu\nu\rho\sigma} F_{\rho\sigma}/2$. From now on we suppress indices, writing $F \tilde{F}$ in place of $F^{a}_{\mu\nu} \tilde{F}^{a,\mu\nu}$. The anomaly coefficients are 
\begin{equation}
\begin{split}
A_{3} \,\delta_{b c} & = N_{\rm DC} \,{\rm Tr}\left[ \{t_{b},t_{c}\} Q_{a} \right] = 2 N_{\rm DC} \,\delta_{b c}\\[0.2cm]
A_{Y} & = N_{\rm DC} \,{\rm Tr}\left[ \{Y,Y\} Q_{a} \right] = 12 y^{2} N_{\rm DC}\, ,
\end{split}
\end{equation}
where $t_{i}$ and $Y$ are the QCD and hypercharge generators respectively. We can thus identify ${\rm U(1)}_{a}$ with the Peccei-Quinn symmetry and the corresponding NGB $a$ will play the role of the QCD axion. The other spontaneously broken accidental symmetry, $\rm U(1)_{s}$, is free from gauge anomalies and its NGB $s$ is exact, hence massless, at the level of the renormalizable Lagrangian. The appearance of such massless state in the IR can be also deduced from an 't Hooft anomaly matching argument.\footnote{For example, one can think of weakly gauging the group $\rm U(1)_{V}$ generated by the vectorlike charge $Q_V = {\rm diag}(1,0,0,\mathbb{1})$, which is free from gauge anomalies. The mixed $\rm [U(1)_{V}]^{2}U(1)_{s}$ anomaly is reproduced at low energy by $s$.}

At the level of discrete symmetries, one can define the action of charge conjugation $C$ and parity $P$ on the new fields as follows\footnote{This definition corresponds to standard transformation rules for the Dirac field $\Psi$: $\Psi(x) \to -i \gamma^2 \Psi^*(x)$ under $C$ and $\Psi(\vec x,t) \to \gamma^0 \Psi^*(-\vec x,t)$ under $P$.}
\begin{align}
C: \begin{cases} 
\psi_i(x) \leftrightarrow \chi_i(x) \\
{\cal G}^{a}_{\mu}(x) T_a \rightarrow -{\cal G}^{a}_{\mu}(x) T_a^* \, ,
\end{cases}
\qquad
P: \begin{cases}
\psi_i(\vec x, t) \leftrightarrow \chi^c_i(-\vec x,t) \\
{\cal G}^{a}_{\mu}(\vec x, t) \rightarrow \eta_{\mu\mu} {\cal G}^{a}_{\mu}(-\vec x, t) \, ,
\end{cases}
\end{align}
where ${\cal G}^{a}_{\mu}$ denotes the $\SUNDC$ gauge fields, $T_a$ are the corresponding generators, and $\eta_{\mu\mu}$ is equal to $+1$ for $\mu=0$ and $-1$ for $\mu=1,2,3$. Both $C$ and $P$ defined in this way are explicitly broken by the chiral $\UoneD$ gauging, but $CP$ is preserved thanks to the absence of a dark color theta angle. It is also possible to define an alternative charge conjugation symmetry, $C'$, which is exact in the new sector and broken only by the EW interactions of SM fields. We define
\begin{equation}
C':
\begin{cases}
\psi_1 \leftrightarrow \psi_2 \, , \quad \psi_3 \leftrightarrow \psi_4  \, ,
\quad \chi_1 \leftrightarrow \chi_2 \, , \quad \chi_3 \leftrightarrow \chi_4  \\
 A_{\mu}^D \rightarrow -A_{\mu}^D \, , \quad G^{a}_{\mu} \lambda_a \rightarrow -G^{a}_{\mu} \lambda_a^* \, ,
\quad B_{\mu}  \rightarrow - B_{\mu}\, , 
\end{cases}
\end{equation}
whereas the $\SUNDC$ gauge fields do not transform. Under $C'$ the NGBs $s$ and $a$ have charge $-1$ and $+1$ respectively, while they are both odd under the action of $CP$.

After dark confinement the $\UoneD$ gauge boson acquires a mass
\begin{equation}
m_{\gamma_{D}} = 2 (1-q) e_{D} f_{\rm DC}\, .
\end{equation}
The pseudo NGBs, whose corresponding global symmetries are explicitly broken by the weak gauging, acquire a mass from loops of SM and/or $\UoneD$ gauge bosons of order $\delta m^2 \sim (\alpha_i/4 \pi) \Lambda_{DC}^2$, where the couplings are evaluated at $\Lambda_{\rm DC}$. Other baryonic and mesonic resonances of the strong dynamics have mass of order $\Lambda_{\rm DC}$.  If $\Lambda_{\rm DC}$ is identified with the Peccei-Quinn scale, all these states are extremely heavy; for values of the $\UoneD$ coupling $\alpha_{D}\gtrsim 10^{-2}$ and $q \sim O(1)$, their masses are  always larger than $10^{8} \, \rm GeV$. All resonances can be classified according to their transformation rules under the unbroken vectorial ${\rm U}(1)$ symmetries and some of them are accidentally stable. They are however out of the reach of current experiments and can be neglected if their cosmological abundance is zero.

At energies below the PQ scale one can integrate out all these heavy states and write an effective theory describing the relevant degrees of freedom, which are the NGBs $a$ and~$s$, and the SM fields. Since the generators $Q_{a}$ and $Q_{s}$ commute, the effective theory is simply that of two shift-symmetric scalar fields, plus terms capturing the anomalies and SM interactions. At quadratic order in the derivatives, and defining the axion decay constant~$f_{a}$ as customary in axion physics, the effective lagrangian at the PQ matching scale is:
\begin{equation}\label{eq:lag}
\mathcal{L}_{\rm eff} = \dfrac{1}{2} (\partial_{\mu} a)^{2} +\dfrac{1}{2} (\partial_{\mu} s)^{2} + \dfrac{\alpha_{3}}{8\pi} \dfrac{a}{f_{a}} \, G \tilde{G}+ c_B  \dfrac{\alpha_{Y}}{8\pi} \, \dfrac{a}{f_{a}} \, B \tilde{B},
\end{equation}
with
\begin{equation}
f_{a} =  \frac{f_{\rm DC}}{2A_3} = \dfrac{f_{\rm DC}}{4 N_{\rm DC}} \, , \qquad c_B = 6y^{2}\, .
\end{equation}
From the latter equation we estimate $\Lambda_{\rm DC} \sim 4\pi f_{\rm DC}/\sqrt{N_{\rm DC}} = 16\pi \sqrt{N_{\rm DC}}  f_a$ in this model.\footnote{Notice that the value of the PQ decay constant $f_{\rm DC}$ depends on how it is defined. A redefinition of the PQ charges, for example, also changes $f_{\rm DC}$ if the latter is defined in terms of the matrix element of the PQ current. The dark color scale $\Lambda_{\rm DC}$ and the axion decay constant $f_a$ set, respectively, the mass of the dark hadrons and the axion coupling to photons. As such, they do not depend on a redefinition of the PQ charges. This means that they are related to $f_{\rm DC}$ in a way that depends on the convention chosen. The naive estimate $\Lambda_{\rm DC} \sim 4\pi f_{\rm DC}/\sqrt{N_{\rm DC}}$ does not take this dependence into account and is thus necessarily a very rough approximation.} Knowing that the number of dark flavors is $N_{f}=8$ and the effective number of additional QCD flavors is $\Delta N_{f}^{(\rm QCD)}= 2N_{\rm DC}$, we can derive the constraints implied by the requirements of dark confinement and perturbativity of SM couplings. We demand that the theory remains perturbative for values of $f_a$ as low as $10^9\,$GeV.
Equations~(\ref{eq:dc_coupling_conf}) and~(\ref{eq:qcd_coupling})  then imply the bound on the number of dark colors
\begin{equation}
\label{eq:boundonNDC}
3 \leq N_{\rm DC} \leq 13\, ,
\end{equation}
where $N_{\rm DC}=3$ is at the edge of the conformal window~\cite{DeGrand:2015zxa}.
Furthermore, Eq.~(\ref{eq:Y_coupling}) gives the bound $y^{2}\lesssim 0.1$ for $N_{\rm DC}$ in the range of Eq.~(\ref{eq:boundonNDC}).

As we discuss in detail in Appendix~\ref{appendix:discrete}, in composite axion models based on a confining vector-like group the number of distinct vacua, also known as the domain wall number $N_{\rm{DW}}$, is equal to the anomaly coefficient computed in the basis where all the PQ charges are coprime integers. In our notation,
\begin{equation}
N_{\rm{DW}} = \vert A_3 \vert = 2 \, N_{\rm{DC}}.
\end{equation}
The UV contribution to the coupling of the axion to photons can be characterized, as usual, by the $E/N$ ratio and its value is controlled in our model by the arbitrary hypercharge $y$; we find $E/N = c_B = 6y^2$.
Higher-derivative interactions, such as $B^{\mu\nu}\partial_\mu s\partial_\nu a$, $(\partial_{\mu} s)^{2} B_{\rho\sigma}B^{\rho\sigma}$, $(\partial_{\mu}a)^{2} (\partial_{\nu} s)^{2}$ etc., are suppressed by higher powers of $f_{\rm DC}$ and can be safely neglected at the energies we are considering. 
By matching the renormalizable UV dynamics to the effective theory at $\Lambda_{\rm DC}$, no derivative operator of the form $\partial_{\mu} a j^{\mu}$, where $j^\mu$ is an axial current made of SM fermions, is generated at leading order in the SM couplings. This is because SM fermions communicate with the new fermions only through SM gauge fields. The same feature characterizes many models of composite axions. Of course, since $\partial_\mu j^\mu$ mixes with $G\tilde G$ and $B\tilde B$ under the renormalization group, operators $\partial_{\mu} a j^{\mu}$ are generated through loop diagrams with one insertion of the anomalous couplings in Eq.~(\ref{eq:lag}). Furthermore, non-renormalizable operators of the form
\begin{equation}
\label{eq:fourfermion}
\bar{\Psi} \gamma_{\mu} \gamma^5 \, \Psi \bar{\Psi}_{SM} \gamma^{\mu} \gamma^5 \Psi_{SM}
\end{equation}
can also generate $\partial_{\mu} a j^{\mu}$, as well as $\partial_{\mu} s j^{\mu}$, at the matching scale if present in the UV theory.

Since the Peccei-Quinn symmetry -- identified with ${\rm U}(1)_{a}$ -- is accidental, it is important to determine at which level it can be broken by higher-dimensional operators. The lowest dimensional PQ-breaking operators consistent with Lorentz and gauge invariance have scaling dimension 6 and are of the form $\psi_{1}\psi_{2}\chi_{1}\chi_{2}$ and $\psi_{3}\psi_{4}\chi_{3}\chi_{4}$. These operators are gauge singlets for arbitrary values of $q$ and carry a non-vanishing Peccei-Quinn charge. Moreover, they have
the appropriate flavour structure to generate a potential for the axion, according to the criterion of Section~\ref{sec:spurion}.

\subsection{A Minimal $ SU(5)$ GUT model}
\label{sec:GUT}

The existence of dimension-6 PQ-breaking operators should prompt one to worry that additional UV dynamics, in particular Grand Unification, may spoil the axion solution. We shall show that it is actually easy to embed the QCD model in a GUT theory in such a way that ${\rm U}(1)_{a}$ is an accidental symmetry of the unified dynamics below the Planck scale. As long as gravitational effects are exponentially suppressed, the theory thus provides a solution to the strong CP problem. In the following we will assume that the GUT group is spontaneously broken at energies higher than the dark confinement scales. This is mostly motivated by the fact that $f_a \leq 1.2\times 10^{16}\,$GeV is required (according to an estimate based on the results of Ref.~\cite{Hebecker:2018ofv}) to protect the axion from too large non-perturbative gravitational effects, and GUT scales smaller than $\sim 10^{16}\,$GeV are generically at odds with proton decay.

Extending the theory is also motivated by its chiral nature: if $M_{\rm GUT}>\Lambda_{\rm DC}$ and the GUT dynamics does not break $\SU(N_{\rm DC})\times {\rm U}(1)_{\rm D}$, as we are assuming, we expect dark fermions charged under the SM to come in complete GUT representations. For the case of (non-supersymmetric) Georgi-Glashow (GG) unification~\cite{Georgi:1974sy}, this can be realized by the straightforward embedding of the color triplets into fundamentals of $\SUfiveGUT$, as shown in Table~\ref{tab:model}. 
With this choice, the number of $\SUNDC$ flavors is $N_{f}=12$, while the effective number of additional QCD flavors is unchanged. As reported in Eq.~(\ref{eq:fa2}), the relation between $f_a$ and $f_{\rm DC}$ is also unchanged. The conditions of confinement and perturbativity, Eqs.~\eqref{eq:dc_coupling_conf} and~\eqref{eq:qcd_coupling}, then imply on $N_{\rm DC}$ the same constraint of Eq.~(\ref{eq:boundonNDC}) as in the QCD model.
As before, the renormalizable lagrangian includes only kinetic terms: no Yukawa terms with GUT scalars are allowed by gauge and Lorentz invariance if these fields do not carry $\UoneD$ charges, as we will assume in the following. Since all fermions transform in complete GUT multiplets, the one-loop differential running of the gauge couplings is the same as in the SM, leaving the accuracy to which unification is realized in the minimal GG model unaltered. This is no longer true at two-loop level, since the splitting among the radiatively generated masses of the pseudo NGBs and threshold corrections can change the differential running. A non-minimal GUT sector with additional scalars or fermions can improve the unification without interfering with the present set-up. 

While most of the qualitative features of the GUT theory are carried over from the simpler QCD model, the additional fields introduce some differences at low energy. We assume that the GUT group is broken at scales higher than $\Lambda_{DC}$, hence the symmetry breaking pattern of Eq.~(\ref{eq:sb1}) is enlarged to
\begin{equation}\label{eq:sb2}
{\rm U}(1)_L^{6} \times {\rm U}(1)_R^{6} \longrightarrow {\rm U}(1)_V^{6}\, ,
\end{equation}
resulting in two additional pseudo NGBs. These correspond to the two SM singlets contained in $\UoneD$-neutral adjoints of $\SUfiveGUT$ in the spectrum of NGBs. We will denote them by $\tilde{a}$ and $\tilde{s}$ and indicate with ${\rm U}(1)_{\tilde a}$ and ${\rm U}(1)_{\tilde s}$ the associated spontaneously broken symmetries. Both ${\rm U}(1)_{\tilde a}$ and ${\rm U}(1)_{\tilde s}$ are explicitly broken by the GUT dynamics, and a mass for $\tilde{a}$ and $\tilde{s}$ of order
\begin{equation}
\label{eq:masNGBstilde}
m_{\tilde a,\tilde s}^2 \sim  \frac{g^2_{\rm GUT}}{16 \pi^2} \frac{\Lambda_{\rm DC}^4}{M^2_{\rm GUT}}
\end{equation}
is generated radiatively by loops of GUT gauge bosons. The same effect arises in the low energy theory from dimension-6 operators that are generated at $M_{\rm GUT}$ by integrating out GUT states. We define the generators so that the ${\rm U}(1)_{\tilde a}$ group is anomalous under $\SU(3)_c\times\SU(2)_{EW}\times \U(1)_Y$, hence $\tilde{a}$ is a heavy axion-like particle, while ${\rm U}(1)_{\tilde s}$ is free from gauge anomalies. Under $CP$ $\tilde{a}$ and $\tilde{s}$ are both odd, while under $C'$ they have charge $+1$ and $-1$ respectively. Since $C'$ is an exact symmetry in the dark sector, the mass matrix of $\tilde s$ and $\tilde a$ is diagonal up to higher-order $C'$-breaking effects. A mixing between $\tilde s$ and $\tilde a$ can only arise through higher order loops involving SM fermions.

According to Eq.~(\ref{eq:masNGBstilde}), $\tilde a$ and $\tilde s$ have masses that range from $\sim 10^4\,$GeV to $\sim 10^{8}\,$GeV when $\Lambda_{\rm DC}$ varies from $10^{11}\,$GeV to $10^{13}\,$GeV.
Their only interactions come from non-renormalizable operators and are too feeble to give appreciable signals in laboratory experiments.
Similarly to the axion, both $\tilde a$ and $\tilde s$ can be very long-lived: $\tilde a$ can decay to two photons through the anomaly, with decay width 
\begin{equation}
\label{eq:decayrateatilde}
\Gamma_{\tilde{a}\to \gamma\gamma} \sim \frac{\alpha_{em}^2}{2(4 \pi)^3} \frac{m^3_{\tilde{a}}}{f_{\rm DC}^2}\, ,
\end{equation}
while $\tilde s$ has an even smaller decay width that can be induced only by a mixing with $\tilde s$ generated by higher-order $C'$-breaking loops. The metastability of $\tilde a$ and $\tilde s$ can have an impact on cosmology, as discussed in Section~\ref{sec:phenomenology}.

Apart from $\tilde{a}$ and $\tilde{s}$, the low-energy limit of the GUT theory is also characterized by the presence of the two light NGBs $a$ and $s$. The generalization of Eq.~(\ref{eq:gen}) is straightforward,
\begin{equation}\label{eq:gen_GUT}
Q_{s} = {\rm diag}(5,-5,-\mathbb{1},\mathbb{1}),
\qquad  
Q_{a} = {\rm diag}(-5,-5,\mathbb{1},\mathbb{1}) \, ,
\end{equation}
and the same properties for $a$ and $s$ are derived as for the QCD model. At energies of order $\Lambda_{\rm DC}$ the theory can be matched to an effective Lagrangian for the four SM-singlet NGBs. This is similar to that of Eq.~(\ref{eq:lag}), with $f_a$ and $N_{DW}$ given again by
 \begin{equation} \label{eq:fa2}
f_a = \dfrac{ f_{\rm DC} }{4 N_{DC}}\qquad \quad N_{DW} =2 N_{DC} ,
\end{equation}
but with two additions. First, there is an anomalous coupling of the axion also to $SU(2)_{EW}$ gauge fields and, as a consequence of the $SU(5)$ invariance, the axion-photon coupling takes the standard value for a model with complete $\SUfiveGUT$ representations, i.e. $E/N = 8/3$. Second, four-fermion operators of the same form as in Eq.~(\ref{eq:fourfermion}) are generated at $M_{\rm GUT}$ by the tree-level exchange of heavy gauge bosons; at the scale $\Lambda_{\rm DC}$, they in turn generate derivative interactions of the form $c_j \partial_{\mu}a j^{\mu}/f_a$ with coefficients
\begin{equation}
c_j \sim \dfrac{g_{\rm GUT}^2f_a^2}{M^2_{\rm{GUT}} }\, .
\end{equation}
For $g_{\rm{GUT}}\lesssim 0.1$ and $\Lambda_{\rm{DC}} < M_{\rm{GUT}}$, the estimated value of $c_j$ is always smaller than $10^{-4}$.

\subsection{Generic Models with $n_f=2$}
\label{sec:nf2_class}

We now generalize the classification to the case of arbitrary $\GSM$ representations $r_{i}= \sum_{\alpha} r_{i}^{(\alpha)}$, with $r_{i}^{(\alpha)}$ irreducible. To this aim, it is convenient to introduce the following short-hand notation: 
\begin{equation}
d_{i}= {\rm dim}(r_i) = \sum_{\alpha} {\rm dim}(r_{i}^{(\alpha)})\, , \qquad T_{i}= \sum_{\alpha} T(r_{i}^{(\alpha)}) \, .
\end{equation}
The choice of $\UoneD$ charges is restricted by the requirement of gauge anomaly cancellation. Linearly independent, non-trivial constraints come from the anomalies $[\GSM]^2\UoneD$, $[\SUNDC]^{2}\UoneD$ and $[\UoneD]^{3}$; following the notation of Eq.~\eqref{eq:genmod}, the corresponding conditions are:
\begin{equation}
\label{eq:anomaly1}
\begin{cases}
(p_{1}+q_{1})T_{1} + (p_{2}+q_{2})T_{2}=0, \\
(p_{1}+q_{1})d_{1} + (p_{2}+q_{2})d_{2}=0, \\
(p_{1}^{3}+q_{1}^{3})d_{1} + (p_{2}^{3}+q_{2}^{3})d_{2}=0\, .
\end{cases}
\end{equation}
Combining the first two equations we obtain
\begin{equation}
\label{eq:anomaly2}
(p_{1}+q_{1}) \left(d_{2}T_{1}- d_{1} T_{2} \right) =0\, .
\end{equation}
The solution $p_{1}=-q_{1}$ implies $p_{2}=-q_{2}$ and leads to a vector-like charge assignment, which we exclude. We are thus interested in choices of representations such that $T_{1} d_{2}=T_{2} d_{1}$.

One class of solutions is obtained for $d_{1}=d_{2}, T_{1}=T_{2}$. This is the only possibility if the choice of the $r_{i}$ is restricted to irreducible representations of $\SU(N)$ with rank 2 or lower, as done in Ref.~\cite{Contino:2020god} to maintain perturbativity up to the Planck scale. In that case, the possible charge assignments are the same as those of Ref.~\cite{Contino:2020god}, which lead to Type I, II and III theories defined at the beginning of this section.
For reducible $r_{i}$, new solutions of Eqs.~(\ref{eq:anomaly1}) and (\ref{eq:anomaly2}) exist with $d_{1} \neq d_{2}$; in the case of a chiral assignment of charges and for given arbitrary $q_{1}$ and $p_{1}$, we find
\begin{equation}
\label{eq:nf2_charges}
\begin{split}
q_{2}= - \dfrac{1}{2} \dfrac{d_{1}}{d_{2}} (p_{1}+q_{1}) \pm \dfrac{\sqrt{12 d_{2}^{4} (p_{1}^{2}-p_{1} q_{1}+ q_{1}^{2})- 3 d_{1}^{2} d_{2}^{2}(p_{1}+q_{1})^{2}}}{6 d_{2}^{2}},
\\[0.2cm]
p_{2}= - \dfrac{1}{2} \dfrac{d_{1}}{d_{2}} (p_{1}+q_{1}) \mp \dfrac{\sqrt{12 d_{2}^{4} (p_{1}^{2}-p_{1} q_{1}+ q_{1}^{2})- 3 d_{1}^{2} d_{2}^{2}(p_{1}+q_{1})^{2}}}{6 d_{2}^{2}}.
\end{split}
\end{equation}
It is always possible to redefine the $\UoneD$ coupling constant in such a way that one of the charges is normalized to one, \textit{e.g.} $q_{1}=1$.

The possible choices of $\GSM$ representations $r_{1}, r_{2}$ are narrowed by the confinement and perturbativity constraints of Section \ref{sec:pt}. As in the case of the minimal QCD and GUT models, we require that there exist a non-vanishing range of $N_{\rm DC}$ in which 
the dark color group confines and the SM couplings remain perturbative for values of $f_a$ as low as $10^9\,$GeV. This criterion selects a sufficiently large and representative set of models, as we will see. Furthermore, when imposing Eq.~(\ref{eq:qcd_coupling}) we assume a relation $\Lambda_{\rm DC} = 100\, f_a$ between the dark confinement scale and the axion decay constant.
This makes it possible to simplify the analysis and it is a good approximation in all the models considered. A thorough and more accurate analysis would require deriving for each model the exact relation between $f_{\rm DC}$ and $f_a$ (by determining the embedding of $\UonePQ$ in the global symmetry group) and varying $f_a$ in the whole allowed range. Some models with larger ratios $f_{\rm DC}/f_a$ that have been excluded by our selection could be reconsidered in this way. We leave such analysis of $n_f=2$ models for future work and will perform it only for the $n_f=3$ models of Section~\ref{sec:nf3}. Besides passing the requirements of confinement and perturbativity, viable models must have at least two fragments with different Dynkin indices in order to have an axion, \textit{i.e.} a NGB with a non-zero anomalous coupling to QCD gauge fields.
The models satisfying these conditions, and the dimension of the corresponding representations, are listed in Table~\ref{tab:nf2_rep}.
\begin{table}
\centering
\begin{tabular}{cl|cc|cc|c|c}
& $\GSM$  & $r_1$ & $r_2$ & $d_1$ & $d_2$ & & $\Delta^{\rm max}_{\cancel{PQ}}$ \\[0.1cm]
\hline
& $\SU (3)_{\rm c}$ & $(\mathbf{3} \oplus m \, \mathbf{1})$ & $(\mathbf{3} \oplus m \, \mathbf{1})$ & $3+m$ & $3+m$ & $1\leq m \leq 23$ & 6\\
& & $(\mathbf{3} \oplus m \, \mathbf{1})$ & $2\, (\mathbf{3} \oplus m \, \mathbf{1})$ & $3+m$ & $2(3+m)$ & $1\leq m \leq 8$ & 9\\
& & $(\mathbf{3} \oplus m \, \mathbf{1})$ & $3\, (\mathbf{3} \oplus m \, \mathbf{1})$ & $3+m$ & $3(3+m)$ & $1\leq m \leq 3$ & 12\\
  & & $2\, (\mathbf{3} \oplus m \, \mathbf{1})$ & $2\, (\mathbf{3} \oplus m \, \mathbf{1})$ & $2(3+m)$ & $2(3+m)$ & $1\leq m \leq 3$ & 6\\
  & & $(\mathbf{3} \oplus \mathbf{1})$ & $4\, (\mathbf{3} \oplus \mathbf{1})$ & $4$ & $16$ & & 15\\
  & & $2\, (\mathbf{3} \oplus \mathbf{1})$ & $3\, (\mathbf{3} \oplus \mathbf{1})$ & $8$ & $12$ & & 15\\
\hline
\hline
& $\rm SU(5)_{GUT}$  & $(\mathbf{5} \oplus m \, \mathbf{1})$ & $(\mathbf{5} \oplus m \, \mathbf{1})$ & $5+m$ & $5+m$ & $1\leq m \leq 21$ & 6\\
  & & $(\mathbf{5} \oplus m \, \mathbf{1})$ & $2\, (\mathbf{5} \oplus m \, \mathbf{1})$ & $3+m$ & $2(3+m)$ & $1\leq m \leq 6$ & 9\\
  & & $(\mathbf{5} \oplus \mathbf{1})$ & $3\, (\mathbf{5} \oplus \mathbf{1})$ & $6$ & $18$ & & 12\\
  & & $2\, (\mathbf{5} \oplus \mathbf{1})$ & $2\, (\mathbf{5} \oplus \mathbf{1})$ & $12$ & $12$ & & 6 \\
& & $\mathbf{5}$ & $(\mathbf{10} \oplus 5 \, \mathbf{1})$ & $5$ & $15$ & & 12\\
\end{tabular}
\caption{\it Choices of representations for models with $n_{f}=2$. The full list of possibilities contains also theories that are obtained by replacing any of the irreducible representations in this table with its conjugate. In the case of the $\SU (3)_{\rm c}$ theories listed in the upper panel, electroweak quantum numbers are not specified but can be assigned freely as long as the conditions from cancellation of gauge anomalies and perturbativity of gauge couplings are satisfied.}
\label{tab:nf2_rep}
\end{table}

We still need to assess to what level the PQ symmetry is protected in these theories. From Eq.~\eqref{eq:anomaly1} it follows that an operator of the form 
\begin{equation}
(\psi_{1}\chi_{1})^{d_{1}} (\psi_{2}\chi_{2})^{d_{2}}
\end{equation}
is always a gauge singlet, since $d_{1}(p_{1}+q_{1})+d_{2}(p_{2}+q_{2})=0$ and $d_{1},d_{2}$ are positive integers. If ${\rm gcd}(d_{1},d_{2})>1$, where ${\rm gcd}(d_{1},d_{2})$ denotes the greatest common divisor of $d_1$ and $d_2$, it is possible to reduce the dimension of the operator by sending $d_{1}\rightarrow d_{1}/{\rm gcd}(d_{1},d_{2})$ and $d_{2}\rightarrow d_{2}/{\rm gcd}(d_{1},d_{2})$. This class of operators has the right form to violate the PQ global symmetries associated to the models of Table~\ref{tab:nf2_rep} and generate a potential for the axion according to the criterion of Section~\ref{sec:spurion}. The PQ-violating dimension $\Delta_{\cancel{PQ}}$, as defined in Sec.~\ref{sec:spurion}, has an upper bound $\Delta_{\cancel{PQ}} \leq \Delta^{\rm max}_{\cancel{PQ}} = 3 (d_{1}+d_{2})/{\rm gcd}(d_{1},d_{2})$.
It follows that models with $d_{1}=d_{2}$, among which are those analyzed in the previous sections, generically have dangerous dimension-6 operators and cannot solve the axion quality problem.
Models with $\Delta^{\rm max}_{\cancel{PQ}} \geq 9$ have instead a chance to solve the quality problem, for suitable values of~$f_a$, depending on the actual value of $\Delta_{\cancel{PQ}}$. Determining the latter requires specifying the $\UoneD$ charges and the embedding of $\UonePQ$ into the global symmetry group. We leave such analysis for future work~\cite{Podo:2022gyj}.

\section{Models with $n_{f}=3$} 
\label{sec:nf3}

In Section \ref{sec:mb} we showed that models featuring a composite axion in the low-energy spectrum are those with at least three irreducible representations. Here we analyze models with exactly three irreducible representations, hence with $n_{f}=\Nirr = 3$. We will show that in this case, for appropriate choices of the $\UoneD$ charges, it is possible to have a sufficiently protected accidental Peccei-Quinn symmetry and thus obtain candidate solutions to the axion quality problem.

As for $n_{f}=2$, we make no a priori assumptions on the SM representations of the dark fermions. We classify models of the form detailed in Table~\ref{tab:model2}. 
\begin{table}
\centering
\begin{tabular}{lccc|c}
& $\SUNDC$ & $\UoneD$ & $\GSM$ & $\UonePQ$ \\
\cline{1-5} 
\rule{0pt}{2.4ex}$\psi_1$ & $\square$ &  $p_{1}$ & $r_1$& $\alpha$ \\
$\psi_2$ & $\square$ &  $p_{2}$ & $r_2$& $\beta$ \\         
$\psi_3$ & $\square$ &  $p_{3}$ & $r_3$ & $\gamma$\\
\cline{1-5}        
\rule{0pt}{2.4ex}$\chi_1$ & $\bar{\square}$ &  $q_{1}$  & $\bar{r}_1$& $\alpha$ \\        
$\chi_2$ & $\bar{\square}$ &  $q_{2}$  & $\bar{r}_2$& $\beta$ \\ 
$\chi_3$ & $\bar{\square}$ &  $q_{3}$   & $\bar{r}_3$ &  $\gamma$\\
\end{tabular}
\caption{\it Composite axion model with $n_f=3$. The representations $r_1,r_2,r_3$ are assumed to be irreducible, and $G_{\rm SM}$ is either the SM gauge group or one of its GUT extensions. The $\UoneD$ charges are defined so that $p_i \not = p_j$, $q_i \not = q_j$ for $i\not = j$  and can be normalized to be all integer numbers.}
\label{tab:model2}
\vspace{0.5cm}
\end{table}
The conditions for the cancellation of $[\GSM]^2\UoneD$, $[\SUNDC]^2\UoneD$ and $[\UoneD]^3$ gauge anomalies are:
\begin{equation}\label{eq:an3f}
\begin{cases}
(p_{1}+q_{1})T_1+(p_{2}+q_{2})T_2+(p_{3}+q_{3})T_3=0,\\
(p_{1}+q_{1})d_1+(p_{2}+q_{2})d_2+(p_{3}+q_{3})d_3=0,\\
(p_{1}^{3}+q_{1}^{3})d_1+(p_{2}^{3}+q_{2}^{3})d_2+(p_{3}^{3}+q_{3}^{3})d_3=0.\\
\end{cases}
\end{equation}
If the representations $r_1,r_2,r_3$ are all equal, then the accidental NGB in the spectrum is not anomalous under SM color and cannot play the role of the QCD axion.  If two representations are equal (up to conjugation), then the system (\ref{eq:an3f}) admits only vector-like solutions. Indeed, in that case from the first two equations of Eq.~(\ref{eq:an3f}) it follows $(p_i+q_i) (T_i d_j - T_j d_i)=0$, where $r_i$ is the representation which differs from the other two. The relation $T_i d_j = T_j d_i$ is not satisfied by any pair of representations of $\SU(N)$ with rank~2 or lower, while higher-dimensional representations are excluded by the perturbativity constraint. This implies $p_i = -q_i$, hence vector-like representations. Therefore, we are led to consider models where the three irreducible representations are all different and inequivalent under complex conjugation.

The choices of representations are restricted by the conditions of Section~\ref{sec:pt}.
We enforce the constraint of Eq.~(\ref{eq:qcd_coupling}) by estimating $\Lambda_{\rm DC} = 4\pi f_{\rm DC}/\sqrt{N_{\rm DC}}$ and computing the ratio $f_{\rm DC}/f_a$ in each model. Together with Eq.~(\ref{eq:dc_coupling_conf}), this leads to a range of allowed values of $N_{\rm DC}$ and $f_a$. The minimum number of dark colors is determined by Eq.~(\ref{eq:dc_coupling_conf}), from which it follows the minimum value of $f_a$ through Eq.~(\ref{eq:qcd_coupling}). We require $f_a\lesssim 10^{12}\,$GeV, and this in turn implies an upper bound on $N_{\rm DC}$ again through Eq.~(\ref{eq:qcd_coupling}).~\footnote{We could consider larger values of $f_a$, up to $M_{\rm GUT}$. For $f_a > 10^{12}\,$GeV, on the other hand, the axion energy density from misalignment is too large unless one assumes small initial conditions, and furthermore the axion quality problem exacerbates.} According to these criteria, there are only a few possibile representations $r_1,r_2,r_3$ leading to viable models. They are reported in Table \ref{tab:arep}, together with the allowed range of dark colors and the minimum value of $f_a$.
\begin{table}
\centering
\begin{tabular}{cl|ccc|c|c|c}
& $\GSM$  & $r_1$ & $r_2$ & $r_3$ & $\Delta^{\rm max}_{\cancel{PQ}}$ & $N_{\rm DC}$ & $f_a^{\rm min}$\\[0.1cm]
\hline
&$\SUfiveGUT$ & $\mathbf{1}$ & $\mathbf{\bar{5}}$ &  $\mathbf{10}$ & $12$ & $4,5,\dots, 11$ & $4\cdot 10^{8}\,$GeV \\
  & & $\mathbf{1}$ & $\mathbf{\bar 5}$ &  $\mathbf{15}$ & $15$& $6,7$ & $10^{11}\,$GeV \\
  & & $\mathbf{1}$ & $\mathbf{10}$ &  $\mathbf{15}$ & $18$ & $7$ & $10^{12}\,$GeV \\
\hline
\hline
& $\SU(3)_{c}$ & $\mathbf{1}$ & $\mathbf{3}$ &  $\mathbf{6}$ & $12$ & $3,4,\dots,9$ & $4\cdot 10^{8}\,$GeV \\
&  &  $\mathbf{8}$ & $\mathbf{3}$ & $\mathbf{6}$ &   $15$ & $5$ & $5\cdot 10^{11}\,$GeV \\
  &  & $\mathbf{1}$ & $\mathbf{3}$ &  $\mathbf{8}$ & $15$ & $3,4,\dots,7$ & $10^{9}\,$GeV \\
&  & $\mathbf{1}$ & $\mathbf{6}$ &  $\mathbf{8}$ &  $12$ & $4,5$ & $10^{11}\,$GeV \\
\end{tabular}
\caption{\it Choices of the representations defined in Table \ref{tab:model2} allowed by the constraints of Section \ref{sec:pt}. Additional viable models are obtained by conjugating the representation $r_3$ (or equivalently $r_2$), except for models in the last two rows. The last two columns report, respectively, the allowed values of $N_{\rm DC}$ and the smallest allowed value of~$f_a$.}
\label{tab:arep}
\vspace{-0.5cm}
\end{table}
Starting from any of the models of this table, except the last two, one can obtain another viable model by conjugating the representation $r_3$ (or equivalently $r_2$); such new theory has a different spectrum of heavy resonances but the same low-energy axion phenomenology compared to its parent theory.

Computing the minimum PQ-violating dimension $\Delta_{\cancel{PQ}}$ defined in Sec.~\ref{sec:spurion} for all the models of Tab.~\ref{tab:arep} is beyond the scope of this work. However, similarly to what we have seen for $n_f=2$ models, it is possible to set an upper bound on $\Delta_{\cancel{PQ}}$ by exploiting the conditions of anomaly cancellation, Eq~(\ref{eq:an3f}). From the first two equations, in particular, it follows that operators of the form
\begin{equation}
\prod_{i=1}^3 \left(\psi_i\chi_i\right)^{\kappa_i} \left(\psi_i^*\chi_i^*\right)^{\bar\kappa_i} 
\end{equation}
are neutral under $\UoneD$ provided that~\footnote{Here we use the fact that $p_i \not = -q_i$ and $T_i d_j \not = T_j d_i$ for any $i,j$ as discussed earlier.}
\begin{equation}
\label{eq:deltamaxcondition}
(\kappa_1-\bar \kappa_1) (T_2 d_3 - T_3 d_2) + (\kappa_2-\bar \kappa_2) (T_3 d_1 - T_1 d_3) + (\kappa_3-\bar \kappa_3) (T_1 d_2 - T_2 d_1) = 0\, .
\end{equation}
These operators are invariant also under $\SUNDC \times \GSM$, manifestly violate $\UonePQ$, and are of the right form to generate an axion potential through a single insertion. Their dimensionality is equal to $3\sum_i (\kappa_i + \bar\kappa_i)$.
Therefore, for a given model of Tab.~\ref{tab:arep}, there exists an upper bound on the dimension of the PQ-violating operators given by $\Delta^{\rm max}_{\cancel{PQ}} = 3 \kappa_{\rm min}$, where $\kappa_{\rm min} = \min\{\sum_i (\kappa_i+\bar\kappa_i)\}$ for positive integer $\kappa_i$, $\bar\kappa_i$ satisfying Eq.~(\ref{eq:deltamaxcondition}).
The values of $\Delta^{\rm max}_{\cancel{PQ}}$ are reported in Tab.~\ref{tab:arep} for the selected models.

All the models of Tab.~\ref{tab:arep} can address the quality problem if their value of $\Delta_{\cancel{PQ}}$ equals the upper value $\Delta_{\cancel{PQ}}^{\rm max}$ computed as above. It is in fact plausible that this equality can be achieved with a suitable choice of the $\UoneD$ charges. In the following we shall focus on GUT models with $(r_1,r_2,r_3) = (\mathbf{1}, \mathbf{\bar{5}},  \mathbf{10})$, where $f_a$ can be as low as $4 \cdot 10^8\,$GeV. We will present their relevant properties and show that choices of $\UoneD$ charges exist for which $\Delta_{\cancel{PQ}} = \Delta_{\cancel{PQ}}^{\rm max} = 12$. QCD models with $(r_1,r_2,r_3) = (\mathbf{1}, \mathbf{3},  \mathbf{6})$ will be briefly discussed in Appendix~\ref{appendix:QCD3}.

\subsection{Analysis of $(1,\bar 5, 10)$ models}

Let us consider models of Table~\ref{tab:model2} with representations $(r_1,r_2,r_3) = (\mathbf{1}, \mathbf{\bar{5}},  \mathbf{10})$ or $(\mathbf{1}, \mathbf{5},  \mathbf{10})$ of $\SU(5)$. In this case the number of dark flavors is $N_f = 16$, and the effective number of additional QCD flavors is $\Delta N_f^{\rm (QCD)} = 4 N_{\rm DC}$. The conditions of Eqs.~(\ref{eq:dc_coupling_conf}), (\ref{eq:qcd_coupling}) restrict the number of dark colors to $4 \leq N_{\rm DC} \leq 11$, and values of $f_a$ as small as $4\cdot 10^8\,$GeV are allowed.
The $\UoneD$ charges must be chosen so as to have chiral representations and this implies that no gauge invariant mass terms are possible. However, differently from the $n_f=2$ case, this condition alone does not automatically forbid the existence of other gauge-invariant 2-fermion operators featuring GUT scalars. These operators explicitly violate the Peccei-Quinn symmetry and their impact on the axion solution should be analyzed carefully. Yukawa operators, in particular, are especially dangerous and one naively expects that they lead to a too large correction to the axion potential unless their coefficient is extremely small. To simplify our analysis, in the following we will consider only sets of  $\UoneD$ charges that do not allow any operator with two dark fermions. In practice, this is equivalent to require $q_i \not = - p_j$ for any $i,j$, and $p_i \not = p_j$,  $q_i \not = q_j$ for any $i\not = j$.

From Section~\ref{sec:mb} we know that the number of NGBs that are singlets of $\GSM$ is equal to $(\Nirr-2)$. For $\Nirr =3$, this means that the only light degree of freedom is the axion itself. The Peccei-Quinn generator can be easily shown to be
\begin{equation}
Q_{a}= \text{diag}(5, \mathbb{1}_{5},-\mathbb{1}_{10})\, .
\end{equation}
The corresponding current $J^{\mu}_{a}$ has an anomaly in the background of $\SU(5)$ gauge fields
\begin{equation}
\langle \partial_{\mu} J_{a}^{\mu} \rangle = - \dfrac{\alpha_{5}}{4\pi} A_{5}\, G \tilde{G} \, ,
\qquad A_{5} = - 2N_{\rm DC}\, ,
\end{equation}
where $G_{\mu \nu}$ is the $\SU(5)$ field strength. In this case the axion decay constant and domain wall numbers are
\begin{equation} \label{eq:fa3}
f_a = \frac{f_{\rm DC}}{4 N_{\rm DC}}  \qquad  \quad N_{DW} =2 N_{DC} \, .
\end{equation}

Besides the axion there are no additional SM-singlet NGBs that play the role of $s$ in $n_f=2$ models. On the other hand, there are three pseudo NGBs, whose mass can be estimated as in Eq.~(\ref{eq:masNGBstilde}). Their number can be easily deduced by noticing that the number of irreducible representations increases from three to six when the GUT group is broken to the SM one. In particular, the three pseudo NGBs originate from the $\mathbf{24}$ contained in the product $\mathbf{5} \times \mathbf{\bar{5}}$, and from the $\mathbf{24}$ and $\mathbf{75}$ contained in the product $\mathbf{10} \times \mathbf{\overline{10}}$. We shall denote them by $\varphi_I$, with $I=1,2,3$. In an orthonormal basis (i.e. ${\rm Tr} (T_I T_J )= \delta_{IJ}/2$), the corresponding generators can be written as
\begin{equation}\label{eq:genALP}
\begin{split}
& T_{1} = \frac{1}{2 \sqrt{5}} \, {\rm diag}(0,3 \cdot \mathbb{1}_2,-2\cdot\mathbb{1}_3, 0,0\cdot\mathbb{1}_{3},0 \cdot\mathbb{1}_{6})\\  
& T_{2} = \frac{1}{6} \, {\rm diag}( 0,0\cdot \mathbb{1}_2,0\cdot \mathbb{1}_3,0,-2 \cdot \mathbb{1}_3, \mathbb{1}_6 ) \\
& T_{3} = \frac{1}{6\sqrt{5}} \, {\rm diag}( 0,0 \cdot \mathbb{1}_2,0 \cdot \mathbb{1}_3,-9, \mathbb{1}_3, \mathbb{1}_6 ).\\
\end{split}
\end{equation}
Then, the anomalous couplings to the SM gauge fields can be written as
\begin{equation}\label{eq:SMan}
\mathcal{L} \supset \dfrac{\varphi_I}{f_{\rm DC}} A_{Ii} \dfrac{\alpha_{i}}{8\pi}  \,G_{i}\, \tilde{G}^{i}\, ,
\end{equation}
where $i$ runs over the three simple factors of the SM gauge group $\SU(3)_c\times\SU(2)_{EW}\times \U(1)_Y$, and the anomaly matrix has the form 
\begin{equation}
A_{Ii} = N_{\rm{DC}}
\begin{pmatrix}
-\frac{1}{\sqrt{15}} & \frac{3}{2 \sqrt{15}} & \frac{5}{6\sqrt{15}} \\
0 & \frac{1}{2} & -\frac{5}{6} \\
\frac{1}{2\sqrt{5}} & \frac{1}{2\sqrt{5}}& -\frac{5}{2 \sqrt{5}} 
\end{pmatrix}.
\end{equation}
Since the matrix $A_{Ii}$ has rank~$2$, one might be tempted to conclude that there is a particular combination of the $\varphi_I$ which does not couple to any of the anomalies, and is thus stable. However, the mass generated radiatively by $\SUfiveGUT$ interactions for the two pseudo NGBs contained in the $\mathbf{24}$ is different from that of the pseudo NGB coming from the $\mathbf{75}$, and furthermore the mass matrix is not diagonal in the chosen basis. We explicitly verified this by computing the masses at 1 loop following Ref.~\cite{Contino:2010rs}.
Therefore, the rotation required to decouple one of the pseudo NGBs from the anomalies would induce a mixing between the $\varphi_I$'s,\footnote{We thank Luca Vecchi for bringing this point to our attention.} and we can conclude that all three states can decay through the anomalies. From a qualitative viewpoint, the situation is very similar to that of the $\tilde{a}$ in the previous section.

Finally, $\UoneD$ is spontaneously broken at dark confinement and the dark photon obtains a mass
\begin{equation}
m_{\gamma_D} = \frac{4}{\sqrt{5}} |p_1+q_1| e_D f_{\rm{DC}}.
\end{equation}
To derive this formula we have used the anomaly conditions~(\ref{eq:an3f}).

\subsection{Explicit solutions and axion quality for $(1,\bar 5,10)$ models}
\label{ssc:exso}

To fully specify a model, one must assign $\UoneD$ charges that satisfy the conditions of anomaly cancellation in Eq.~(\ref{eq:an3f}). The choice of charges determines the structure of higher-dimensional operators and therefore the degree to which the quality problem can be solved in a given model. Finding all the integer solutions to the system~(\ref{eq:an3f}) does not appear to be an easy problem and it will be tackled in a separate work~\cite{Podo:2022gyj}. However, some explicit solutions are easy to find.
A particularly interesting family of solutions is given by
\begin{equation}\label{eq:c10}
(p_1,p_2,p_3) = (5,-3,1) \quad \quad \quad (q_1,q_2,q_3) = -q\, (5,-3,1)\ , 
\end{equation}
where the parameter $q$ is an integer number that must be different than 1 to have chiral representations. That of Eq.~(\ref{eq:c10}) turns out to be the only solution of the form $q_i = - q \, p_i$ with $q$ constant. Its existence can be traced back to the possibility of embedding $\SU(5) \times \U(1)$ into $\SO(10)$ with the branching rule\footnote{This embedding is valid only if $(r_1,r_2,r_3)=(\mathbf{1},\mathbf{\bar{5}},\mathbf{10})$, but since the mixed anomaly cancellation conditions are not sensitive to the exchange $\mathbf{5} \leftrightarrow{\mathbf{\bar{5}}}$, the solution exists also for $(r_1,r_2,r_3)=(\mathbf{1},\mathbf{5},\mathbf{10})$.}
\begin{equation}
\mathbf{16} = \mathbf{1}_5 \oplus \mathbf{\bar{5}}_{-3} \oplus \mathbf{10}_1\, .
\end{equation}
With the correct normalization for $\alpha_D$, the two subgroups can be identified with $\SUfiveGUT$ and $\UoneD$, the $\psi_i$ fields can be organized into a single spinorial representation of $\SO(10)$, and the same is true for the $\chi_i$.
Since the $\mathbf{16}$ is anomaly free, anomaly cancellation emerges automatically and is satisfied independently by each of these two sets of fermions. The model can thus be viewed as the low-energy limit of a more minimal theory with an extended unification group $\SO(10)_{\rm L} \times \SO(10)_{\rm R}$, and a structure that is highly reminiscent of the theories discussed in Ref.~\cite{Vecchi:2021shj}. The precise dark fermion content is shown in Tab.~\ref{tab:so10}, from which it is apparent that the doubling of the $\SO(10)$'s is necessary to ensure a chiral structure.\footnote{A single $\SO(10)$ would lead to $q=1$ in Eq.~(\ref{eq:c10}), hence to vectorlike $\UoneD$ charges.}
\begin{table}
\centering
\begin{tabular}{lccc}
& $\SUNDC$ & $\SO(10)_{\rm L}$ & $\SO(10)_{\rm R}$  \\
\cline{1-4} 
\rule{0pt}{2.4ex}$\psi$ & $\square$ &  $\mathbf{16}$ & $\mathbf{1}$ \\
\cline{1-4}        
\rule{0pt}{2.4ex}$\chi$ & $\bar{\square}$ &  $\mathbf{1}$  & $\mathbf{\overline{16}}$ \\        
\end{tabular}
\caption{\it UV completion of the $n_f =3$ model with the choice of  $\UoneD$ charges in Eq.~(\ref{eq:c10}). SM fermions are assumed to be charged under $\SO(10)_L$.}
\label{tab:so10}
\end{table}
As in \cite{Vecchi:2021shj}, all SM generations can be taken to transform as $\mathbf{16}$ of either $\SO(10)$, and without loss of generality we choose $\SO(10)_{\rm L} $. With a suitable choice of GUT scalars, the group $\SO(10)_{\rm L} \times \SO(10)_{\rm R}$ can be broken down to $\SUfiveGUT \times \UoneD$ at an intermediate scale $M_{\rm GUT}'$, where $\SUfiveGUT$ is the diagonal $\SU(5)$ subgroup and $\UoneD$ is a linear combination of the two ${\rm U(1)}$'s contained in $\SO(10)_{\rm L} \times \SO(10)_{\rm R}$. One thus recovers the original model. The Peccei-Quinn symmetry is explicitly violated by the $\SO(10)_{\rm L} \times \SO(10)_{\rm R}$ gauging, and emerges as accidental at low energies. Unfortunately, the appeal of this construction is diminished by the existence of the PQ-violating operators
\begin{equation}
\mathcal{O}_{L} = \psi_1 \psi_2 \big( \psi_3^* \big)^2\, , \quad \quad \quad \mathcal{O}_{R} = \chi_1 \chi_2 \big( \chi_3^* \big)^2
\end{equation}
that can generate a potential for the axion when combined into a double insertion. Their product indeed has the right form, as discussed in Section \ref{sec:spurion}. Since $\mathcal{O}_{L}$, $\mathcal{O}_{R}$ have dimension 6, their double insertion is equivalent to a single insertion of a dimension-8 operator. Therefore, in this model $\Delta_{\cancel{PQ}} \leq 8$ and the axion solution of the strong CP problem is spoiled for any phenomenologically viable value of the Peccei-Quinn scale.

In general, one can perform a numerical scan of the charges lying within a finite interval $|p_i|,|q_i| \leq n_{\rm{max}}$, and identify models with a high quality axion. An intrinsic limitation is given by the existence of higher-dimensional operators that can affect the axion potential for any values of the charges, as a consequence of anomaly cancellation. Solving Eq.~(\ref{eq:deltamaxcondition}) leads to three dimension-12 operators that are gauge-invariant for any choice of the $\UoneD$ charges and have the correct flavor structure to generate an axion potential at low energy:
\begin{equation}
\mathcal{O}_1= \psi_2 \chi_2 \big( \psi_3 \chi_3 \big)^3\, , \quad 
\mathcal{O}_2= \psi_1 \chi_1 \psi_2 \chi_2 \big( \psi_3^* \chi_3^* \big)^2\, , \quad
\mathcal{O}_3= \psi_1 \chi_1 \big(\psi_2 \chi_2\big)^2  \psi_3 \chi_3\, .
\end{equation}
Therefore, $\Delta_{\cancel{PQ}} \leq 12$ in this case, as shown in Tab.~\ref{tab:arep}. We have thus performed a search for models with  $\Delta_{\cancel{PQ}} =12$ (dubbed {\it high-quality} models in the following), which can resolve the quality problem for values of the axion decay constant as high as $10^{12}\,$GeV and explain the whole DM abundance in terms of axions. The result of this search are reported below.

A complete classification depends on the precise representations of the GUT scalars required by the unification dynamics. This is because these fields can help to render some PQ-violating operators gauge invariant, at the price of a slight increase in their dimension. In general, only the following multiple insertions have $d_{\rm eff} < 12$:
\begin{itemize}
\item Two 4-fermion operators, with at most three GUT scalars in their product. 
\item Three 4-fermion operators, with at most one GUT scalar in their product. 
\item One 6-fermion and one 4-fermion operator, with no GUT scalars.
\end{itemize}
In principle these operators might contain SM fermions in flavor-singlet combinations. In practice, $\SUNDC$ invariance and the absence of 2-fermion operators exclude this possibility for 4-fermion operators. One can instead construct gauge invariant 6-fermion operators with four dark fermions and two SM fermions.\footnote{For $N_{DC } = 5$, one can construct a gauge invariant operator with 5 SM fermions and one dark fermion. However, it would need to be paired with another 6-fermion operator to be of the correct form, resulting in an effective dimension $d_{\rm{eff}} =14$.}  However, for representations $r_{1}=\mathbf{1}, r_{2}=\mathbf{\bar{5}}, r_{3}=\mathbf{10}$, it is possible to show that the product of four dark fermions can be always turned into a singlet of $\SU(5)$ by multiplying it with one or two GUT scalars.\footnote{Given any such operator, one can always contract as many of its upper and lower indices as possible with delta tensors and remain with only one type of indices. Their corresponding representation is a product of antisymmetric tensors (each of rank 1 or 2) which always contains a fully antisymmetric tensor. In the case of $\SU(5)$ any antisymmetric tensor can be contracted with at most two fundamentals to give a singlet. The presence of a GUT scalar transforming as a fundamental of $\SU(5)$ is instrumental to embed the Higgs boson.} In this case, the 6-fermion operator can be replaced by an operator with four dark fermions and up to two scalars that has lower dimension and same flavor structure. One can thus focus on operators made of dark fermions only.

Although the classification of dangerous operators depends on the scalar content of the model, it is possible to impose a stronger condition and perform the same analysis without imposing $\SU(5)$ gauge invariance on all operators that feature GUT scalars. This selects a smaller class of
models, dubbed \textit{robust} in the following, that have high quality independently of the choice of GUT scalars, provided these do not carry $\UoneD$ charge.\footnote{In practice, robust models are selected by imposing $\SU(5)$ gauge invariance only on the 6-fermion and 4-fermion operators involved in a double insertion, and on two of the 4-fermion operators involved in a triple insertion.} Besides selecting robust models, we also performed a search for high-quality models with a minimal scalar sector, i.e. GUT scalars transforming as $\mathbf{5}+\mathbf{24}$ of $\SU(5)$.

Table~\ref{tab:hqax} illustrates the result of our numerical study of the parameter space.\footnote{To perform the classification, we have made use of the \textit{Mathematica} package LieART \cite{Feger:2012bs,Feger:2019tvk}.}
\begin{table}
\centering
\begin{tabular}{cl|ccc|ccc|}
&                                    &  & $5\leq N_{\rm{DC}}\leq 11$ &  &  & $N_{\rm{DC}}=4$ &  \\
& $n_{max}$                    & $10$ & $15$ & $20$  & $10$ & $15$ & $20$ \\
\hline
& AC solutions            & $77$ & $189$ &  $341$ & $77$ & $189$ &  $341$   \\
\hline
& HQ axions, $(\mathbf{1},\mathbf{\bar{5}},\mathbf{10})$              & $22$ & $ 68$   &  $150$ & $9$ &  $31$ &  $82$\\
&Robust HQ axions, $(\mathbf{1},\mathbf{\bar{5}},\mathbf{10})$     & $4$   & $ 16$   &  $47$   &  $0$ & $5$ &  $22$ \\
\hline
& HQ axions, $(\mathbf{1},\mathbf{5},\mathbf{10})$              & $14$ & $44$   &  $99$ & $2$ &  $12$ &  $33$\\
& Robust HQ axions, $(\mathbf{1},\mathbf{5},\mathbf{10})$        & $3$ & $16$   &  $36$ & $0$ &  $7$ &  $21$\\ \hline
& No $d\leq 8$ operators & $0$   & $1$      &  $10$   & $0$  & $0$ &  $0$\\
\hline
\end{tabular}
\caption{\it Number of $(\mathbf{1},\mathbf{\bar{5}},\mathbf{10})$ and $(\mathbf{1},\mathbf{5},\mathbf{10})$ GUT models with $\Delta_{\cancel {PQ}}=12$ and integer charges satisfying $|p_i|,|q_i| \leq n_{\rm{max}}$. Models are found through a numerical scan of the parameter space for different values of $n_{\rm{max}}$ and $4\leq N_{\rm{DC}}\leq 11$. The counting is made identifying solutions that are equal up to an overall factor or by exchanging $p_i\leftrightarrow q_i$. The first row indicates the total number of independent solutions of the Anomaly Cancellation (AC) conditions of Eq.~(\ref{eq:an3f}), while the following rows list how many of these solutions yield High Quality (HQ) models, either with a minimal GUT sector or for generic GUT scalar sectors (robust models). The last row shows the number of models with no flavor-violating operators with dimension $d \leq 8$.}
\label{tab:hqax}
\vspace{0.1cm}
\end{table}
Solutions are identified if they only differ by an integer multiplicative factor, or if they can be obtained from one another through the exchange $p_i \leftrightarrow q_i$. The anomaly cancellation conditions alone admit a large number of solutions even for small $n_{\rm{max}}$, and it is precisely thanks to this abundance of solutions that we are able to find examples where the axion quality is protected up to a sufficiently high level.
The classification does not depend on the number of dark colors, except for the special value $N_{\rm DC}=4$. In the latter case it is possible to make $\SUNDC$ singlets by pairing 4 (anti)fundamentals, hence more flavor-violating operators are allowed, implying fewer solutions.
With $n_{\rm{max}} =10$, the following examples are robust in the sense discussed above for $(r_1,r_2,r_3)=(\mathbf{1},\mathbf{\bar{5}},\mathbf{10})$ and $5\leq N_{\rm{DC}}\leq 11$:
\begin{equation}\label{eq:solex}
\begin{split}
&(p_1,p_2,p_3) = (-5,-6,+2) \quad \quad \quad (q_1,q_2,q_3) =  (+10,+3,-1) \\
&(p_1,p_2,p_3) = (+0,-7,+7) \quad \quad \quad (q_1,q_2,q_3) =  (+5,+4,-6) \\
&(p_1,p_2,p_3) =  (+0,+4,-6) \quad \quad \quad (q_1,q_2,q_3) =  (+5,-7,+7) \\
&(p_1,p_2,p_3) =  (+2,+3,-5) \quad \quad \quad (q_1,q_2,q_3) =  (+3,-6,+6) \, ,\\
\end{split}
\end{equation}
and the last three are also robust for $(r_1,r_2,r_3)=(\mathbf{1},\mathbf{5},\mathbf{10})$. The other 18 high-quality solutions are reported in Appendix \ref{appendix:hq}, together with those for $N_{\rm{DC}}=4$.
We stress that we have not found any charge assignment with $n_{\rm{max}} = 20$ where a single insertion of an operator with dimension $d<12$ can generate a potential for the axion: it is the presence of multiple insertions which rules out the excluded models. 
In this sense, the analysis carried out in Section~\ref{sec:spurion} is crucial to correctly identify the high-quality models.

Finally, it interesting to notice that $n_{f}=3$ also allows models with the much stronger property of having no PQ-violating operators (of any form) up to dimension 12. Assuming that invariance under $\SUfiveGUT$ can always be obtained with suitable GUT scalars, there are no such models with $n_{\rm{max}} = 20 $, and the one with the smallest charges that we have found is
\begin{equation}\label{eq:noPQ}
(p_1,p_2,p_3) = (+30,+39,+87) \quad \quad \quad (q_1,q_2,q_3) =  (+55,-90,-70).
\end{equation}

\section{Phenomenology}
\label{sec:phenomenology}

The rich spectrum of composite states predicted by our models in the Peccei-Quinn sector is rather elusive and out of the reach of collider experiments. States with mass of order $\Lambda_{\rm DC}$ are too heavy to be produced, while SM-neutral NGBs have too weak interactions with the SM at low energy. On the other hand, their presence can have an impact on the cosmological evolution and can thus be tested by celestial observations. 
The cosmology of axion models is complex, for a review see for instance Refs.~\cite{Marsh:2015xka, Hui:2021tkt}. A scenario where the scale of inflation $H_{I}$ is larger than the dark confinement (and PQ) scale, or where the PQ symmetry is restored after reheating, appears to be strongly disfavoured in our models. Indeed, in this case the degrees of freedom in the PQ sector would thermalize with the SM bath after reheating, thanks to the QCD interactions of the dark fermions. After confinement, the heavy and accidentally-stable states of the PQ sector (such as dark baryons) would acquire a relic abundance much larger than the observed DM density, in conflict with observations (for a detailed analysis see~\cite{Contino:2020god}). Moreover, as we show in Appendix~\ref{appendix:discrete}, the domain wall number is $N_{\rm{DW}} >1$ in all of the examples, allowing the formation of stable domain wall and string networks. If the PQ phase transition occurs after inflation, they will quickly dominate the energy density of the universe, in conflict with standard cosmology \cite{Zeldovich:1974uw,Sikivie:1982qv}.

We focus, therefore, on the scenario where the PQ symmetry is broken during inflation and the strongly-interacting sector is in its confined phase. In this case, all the accidentally-stable states with mass $\sim \Lambda_{\rm DC}$ are heavier than the scales of inflation and reheating, and are not populated at those temperatures. 
We thus consider only the light dark states, \text{i.e.} the QCD axion $a$ and the other SM-neutral NGBs. The latter include, in general, NGBs that are exact up to non-perturbative gravitational effects and thus almost massless, and heavier pseudo NGBs whose associated symmetry is explicitly broken by GUT corrections. In the following we will analyze models with $n_{f}=2$, where these states are respectively the singlet $s$ and, in GUT models, the $\tilde{a}$ and $\tilde{s}$; GUT models with $n_{f}=3$  and irreducible representations have instead only pseudo NGBs, whose cosmology and phenomenology are similar to those of $\tilde{a}$.

Let us discuss first the massive pseudo NGBs.
The decay rate of Eq.~(\ref{eq:decayrateatilde}) is sufficiently large so that $\tilde{a}$ decays to photons before Big Bang Nucleosynthesis (BBN) for all the phenomenologically allowed values of the PQ scale, $f_{a}\gtrsim 4\cdot 10^8\,$GeV. On the other hand, $\tilde{s}$ is more long-lived, since it decays only through higher-order $C'$-breaking effects. The mass of $\tilde{a}$, $\tilde{s}$ is parametrically suppressed with respect to $f_{a}$ by a factor of order $16\pi^{2}\sqrt{(\alpha_{\rm GUT}/4\pi)} \left(f_{a}/M_{\rm GUT}\right)$, see Eq.~(\ref{eq:masNGBstilde}), and there appear to be two broad scenarios: either \textit{i)} the reheating temperature is smaller than $m_{\tilde a, \tilde s}$, or \textit{ii)} it is in the range $m_{\tilde a, \tilde s} < T_{\rm RH} < f_{a}$. In the first case, the states $\tilde a$ and $\tilde{s}$ are not populated at reheating, the cosmological history is the standard one and the DM abundance can be reproduced by the axion with the usual misalignment mechanism. In the second case, $\tilde a$ and $\tilde{s}$ can be populated at reheating. If $T_{\rm RH} $ is larger than the temperature $T_{\rm fo} \sim m_{\tilde a} (M_{\rm GUT}/M_{\rm Pl}) \sqrt{4\pi}/(\alpha_s^3 \alpha_{\rm GUT}^{1/2})> m_{\tilde a}$ at which processes $gg \leftrightarrow g\tilde a$ go out of equilibrium, $\tilde a$ can reach thermal equilibrium with the SM bath. After an early freeze-out, its energy density will be large and can lead to an epoch of early matter domination, before decaying to photons prior to BBN. In this scenario the abundance of axions given by the misalignment mechanism can be depleted, increasing the value of $f_{a}$ required to match the observed DM abundance for fixed initial misalignment. 
The interactions of $\tilde s$ with the SM bath are more suppressed than those of $\tilde a$, and the region of parameter space where it reaches thermal equilibrium, if any, will be small. When the reheating temperature is higher than their mass but lower than their freeze out temperature, $\tilde a$ and $\tilde s$ have an abundance that depends on the reheating dynamics and
cannot be robustly estimated.
Considering that its decay can happen after BBN, a too large density of $\tilde s$ can be problematic and in conflict with observations.
It would be particularly interesting to explore the possible role of $\tilde{a}$ and $\tilde{s}$ in cosmic inflation and their imprints as primordial non-gaussianities in cosmological correlators, especially because their mass is parametrically separated from $\Lambda_{\rm DC}$ and can be close to the scale of inflation. We leave a more detailed analysis of the role of $\tilde{a}$ and $\tilde{s}$ in cosmology to a future work. In the following, we focus on the first scenario where $T_{\rm RH}$ is smaller than the mass of $\tilde{a}$ and $\tilde{s}$, and give a brief overview of the low-energy phenomenology.

Let us start considering the QCD axion $a$. Its phenomenological properties have been thoroughly explored and are well predicted in terms of the UV parameters by combining chiral perturbation theory and Lattice QCD results~\cite{diCortona:2015ldu,Bonati:2015vqz}. Using the notation and the results of Ref.~\cite{diCortona:2015ldu}, we quote here the predictions in our models for the low-energy parameters relevant for experimental searches. At energies below the QCD confinement scale, and including low-energy QCD contributions, the axion mass, the axion-photon coupling and the axion-nucleon couplings are:
\begin{equation}\label{eq:lag_low}
\begin{split}
&m_{a}= 5.70 (7) \left(\dfrac{10^{12}\, \rm GeV}{f_{a}}\right) \, \rm \mu eV, \\[0.2cm]
& g_{a\gamma\gamma} = \dfrac{\alpha_{em}}{2\pi f_{a}} \left( \dfrac{E}{N}- 1.92 (4) \right), \\[0.2cm]
& c_{p}=-0.47(3), \quad c_{n}=-0.02(3)\, ,
\end{split}
\end{equation}
while the coupling with electrons is generated only through radiative electromagnetic corrections and is suppressed~\cite{Srednicki:1985xd,Chang:1993gm},
$c_{e } \lesssim 3\cdot 10^{-4}$.
The vanishing of the leading-order UV contributions to the axion-fermion couplings in our model gives a sharp prediction for these quantities, common to all the so-called ``hadronic" axion models. Corrections to these predictions from GUT thresholds can be generated, as mentioned in Section~\ref{sec:GUT}, but are suppressed for $\Lambda_{\rm DC}< M_{GUT}$. The low-energy coupling to photons is predicted in the $\SU(5)$ models where $E/N=8/3$, while in QCD models it depends on the hypercharge of dark fermions and is effectively a free parameter.

The cosmological abundance of QCD axions given by the misalignment mechanism can match the observed dark matter abundance for an initial misalignment $\theta_{0}= a_{0}/f_{a}$ of order one and values of the axion decay constant $f_{a} \sim 10^{12}\, \rm GeV$. Larger values of the decay constant require smaller initial misalignment, up to $\theta_{0}\sim 0.01$ for $f_{a} \sim 10^{16}\, \rm GeV$, or a scale of inflation lower than the QCD scale~\cite{Graham:2018jyp,Guth:2018hsa}. Values of $f_{a}$ smaller than $\sim 10^{12}\, \rm GeV$  are allowed and can correspond to the axion accounting for either a subdominant DM component, if $\theta_{0}\simeq 1$, or all of the dark matter, if its initial misalignment is tuned to $\theta_{0}\simeq \pi$.

In the scenario we are considering, $f_{a}> H_{I}$, quantum fluctuations of the axion field during inflation give rise to isocurvature perturbations~\cite{Linde:1985yf}. If the axion accounts for the whole DM density, the current limits on isocurvature perturbations give an upper bound on the scale of inflation $H_{I}\lesssim 10^{9}\, \rm GeV$ for $f_{a}\lesssim 10^{16}\, \rm GeV$, and also a lower bound on the axion decay constant $f_{a} \geq 10^{10}\, \rm GeV$~\cite{diCortona:2015ldu}. The upper bound on $H_I$ implies that in single-field slow-roll inflation the tensor to scalar ratio is extremely small $r \lesssim 2 \times 10^{-6}$. The observation of primordial gravitational waves could therefore falsify our models under the assumption of single-field slow-roll inflation, unless the axion is a subdominant component of the DM abundance.

The $n_f=2$ models of Section~\ref{sec:nf2} predict an additional light NGB, the singlet $s$. This is an exact NGB up to non-perturbative gravitational effects, and has extremely suppressed interactions with the other fields. It behaves effectively as a free massless real scalar field and, as long as its cosmological density is small enough, it can be a harmless new light particle with feeble interactions. Gravitational corrections at the Planck scale can violate the symmetry ${\rm U}(1)_{s}$ and generate a mass $m_s$ such that $\log\left(m_s / M_{\rm Pl}\right) \sim -M_{\rm Pl}/f_a$. For values of the mass in the range $10^{-20}\, {\rm eV} < m_{s} < 10^{-10}\, {\rm eV}$, the existence of this particle may be tested through black hole observations~\cite{Brito:2015oca,Arvanitaki:2016qwi}. An estimate along the lines of Ref.~\cite{Hebecker:2018ofv} seems to suggest $f_{a}\gtrsim 5 \cdot 10^{16}\, \rm GeV$ for this to be the case, with $m_{s}$ exponentially smaller for lower $f_{a}$. Since $f_{a}< 1.2 \cdot 10^{16}\, \rm GeV$ is required in our models to have a high-quality axion, the mass and the abundance of $s$ should be extremely small, according to these estimates. In light of the exponential sensitivity to order one numerical factors, and keeping an open mind on the nature of the UV effects, it could be possible for $s$ to have a mass in the right range to play the role of a fuzzy dark matter component with abundance generated through the misalignment mechanism~\cite{Hui:2016ltb}.

To conclude, both the QCD axion $a$ and the singlet $s$ can have a thermal component, which behaves as dark radiation and can be parametrised by a correction in the effective number of neutrinos $\Delta N_{\nu}^{\rm eff}$. An irreducible component for the QCD axion comes from the thermal production of axions from the SM bath~\cite{Salvio:2013iaa,Arias-Aragon:2020shv,DEramo:2021psx}. Its value depends on the reheating temperature and on $f_{a}$, but it is always bounded by $\Delta N_{\nu}^{\rm eff} \leq 0.026$. Additional contributions can be present, both for $a$ and $s$, depending on the dynamics of reheating or possibly from the decay of a population of primordial black holes~\cite{Hooper:2020evu}. In our treatment these can be seen as initial conditions and do not give rise to sharp predictions, but can be a helpful probe of our models.

\section{Conclusions}
\label{sec:con}

The QCD axion solution to the strong CP problem hinges on the existence of a spontaneously broken Peccei-Quinn symmetry and 
is realized in terms of a very simple EFT at low energies. Despite this apparent simplicity, the PQ symmetry is usually imposed by hand, and explicit models are extremely sensitive to PQ-violating effects in the UV. Global symmetries are ultimately broken by quantum gravity and can be at best approximate. This is especially true for the PQ symmetry, which must be also broken by QCD instantons. Approximate global symmetries emerge at low energy in the effective field theory if they are broken only by irrelevant operators, and in that case they are called accidental. Proton stability and the smallness of neutrino masses can be elegantly explained as consequences of accidental symmetries of the SM. The stability of dark matter could also be the consequence of an accidental symmetry. In all these cases, it is enough to protect the relevant global symmetry up to operators of dimension six to be compatible with current experimental data. Interestingly, breaking effects at the dimension-6 level imply proton decay rates (for a cutoff of order of the GUT scale) and DM decay rates (for a cutoff of order of the Planck scale and DM masses in the $O(100\,\text{TeV})$ range) that are not far from current experimental sensitivities. This means that the paradigm of accidental symmetries could be tested experimentally in the near future. It is hence natural to consider whether the Peccei-Quinn symmetry may also be accidental, that is, emerge at low energy in the effective field theory. While this is an attractive solution, the fact that the axion potential is generated by QCD dynamics at an energy vastly smaller than the cutoff scale, and the existence of stringent bounds from neutron EDM experiments
put the PQ symmetry in a different class compared to the accidental symmetries mentioned above.
In order for the QCD axion solution to work, PQ-breaking operators must have at least dimension 9, while the most attractive scenario where the axion can account for the whole DM abundance requires dimension 12 or larger. This makes the model building of accidental axions much harder than for other SM extensions.

It is clearly possible that the PQ symmetry is not accidental, at least in the sense that it does not emerge accidentally due to a separation of scales in the effective field theory. One can imagine a situation where PQ-breaking effects arise only at the non-perturbative level in the full UV theory, such that any other contribution to the axion potential is much smaller than that from QCD instantons. Models explicitly realizing this scenario have been proposed in the context of string theory. In these constructions the 4D effective action has a non-linearly realised PQ symmetry which acts as a shift symmetry on a fundamental scalar field. The strong CP problem is solved as long as the UV sources of PQ breaking (such as string theory instantons) are small enough, which can be achieved in some of the proposed models. In theories of this kind the axion quality entirely relies on the properties of the UV dynamics and cannot be explained within the effective field theory.

In this work, we have introduced a class of composite axion models where the PQ symmetry is an accidental symmetry of the effective field theory.%

Our constructions are compatible with an \SU(5) unified dynamics at an intermediate scale, although the existence of such a dynamics is not required. The \SU(5) models that we describe in Sections~\ref{sec:nf2} and~\ref{sec:nf3} can be downgraded to purely SM models with straightforward changes, without impairing the solution to the quality problem. Moreover, purely QCD models based on a different choice of representations have been described in Section \ref{sec:qcdm} and in Appendix~\ref{appendix:QCD3}. In all cases, the QCD axion emerges as the pseudo NGB of an accidental $\UonePQ$ global symmetry, and it can be accompanied by other light particles, depending on the specific field content. For $n_{f}=2$ models (both GUT and SM), there is an additional  scalar singlet, lighter than the QCD axion and massless up to UV symmetry breaking effects. Moreover, in the GUT case (for both $n_{f}=2$ and $n_{f}=3$) axion-like particles (ALPs) or additional pseudo NGBs appear, with an intermediate mass scale of order $(g_{\rm GUT}/4\pi) (\Lambda^2_{\rm DC}/M_{\rm GUT})$.
  
Since PQ-violating higher-dimensional operators can spoil the solution of the strong CP problem, it becomes crucial to understand their effects on the low-energy physics in detail, in particular on the axion potential. For a confining $\SUNDC$ gauge group and $\SUNDC \times \GSM $ vectorlike dark fermion representations, we provided in Section~\ref{sec:spurion} simple criteria to discern whether a given set of higher-dimensional operators can generate a potential for the axion (and possibly other NGBs). These are derived from a spurion analysis within the effective field theory as well as using the selection rules associated to the exact global symmetries. Although the form of operators contributing to the potential with a single insertion was already stated by Ref.~\cite{Dobrescu:1996jp}, the inclusion of flavour breaking weak gaugings and the importance of multiple insertions had not been discussed in the previous literature. In our examples, these effects turn out to play a crucial role in excluding otherwise viable models.
Another result of our work is a general analysis of the discrete symmetries associated to $\U(1)_{\rm PQ}$ in composite axion models based on a QCD-like confining gauge group, detailed in Appendix~\ref{appendix:discrete}.

The first class of models we considered are those with two dark flavours ($n_{f}=2$) and generic reducible representations under the SM. They are discussed in Section~\ref{sec:nf2}. The simplest version has the matter content shown in Tab.~\ref{tab:nf2}, while Tab.~\ref{tab:nf2_rep} reports all the assignments compatible with perturbativity of SM gauge couplings and dark color confinement. The $\UoneD$ charges must satisfy the anomaly cancellation conditions but are otherwise left as free parameters in the analysis.
The second class of models, with three irreducible flavours ($n_{f}=3$), is discussed in Section~\ref{sec:nf3}; see Tabs.~\ref{tab:model2} and~\ref{tab:arep} in particular. For both $n_f=2$ and $n_f=3$, we have found several models where the leading PQ-breaking operators have dimension 9 or higher, and that can thus address the axion quality problem.

It is useful to compare the level of PQ symmetry protection achieved in the class of models studied in this work with that of previous works on composite axions.
In the QCD models of Ref.~\cite{Randall:1992ut}, the PQ-violating dimension depends on the number $m$ of colors of the weakly-gauged non-abelian group: $\Delta_{\cancel{PQ}} =3m$. For this reason, a careful analysis of perturbativity of the QCD coupling is crucial to assess the viability of these models as solutions of the quality problem~\cite{Dobrescu:1996jp}.
In the case of simple GUT extensions of the models of Ref.~\cite{Randall:1992ut}, we find that multiple insertions with effective dimension~10 can generate an axion potential, so that $\Delta_{\cancel{PQ}} = \min\{3m, 10\}$, see Appendix~\ref{appendix:Randall}.
The models of Ref.~\cite{Gavela:2018paw} can have an accidental PQ symmetry up to dimension 9, while $\Delta_{\cancel{PQ}} =12$ in the models of Refs.~\cite{Redi:2016esr,Vecchi:2021shj}.
The constructions of Ref.~\cite{Fukuda:2017ylt} can lead to an even higher level of protection, but  do not seem complete, in that gauge anomaly cancellation is postulated through the inclusion of an unspecified set of massless fermions whose charges and low-energy dynamics have not been investigated yet.

The PQ-violating dimension in our models depends on the choice of $\UoneD$ charges. By performing a detailed scan of $n_f =3$ models with GUT representations $(r_1,r_2,r_3) = (\mathbf{1}, \mathbf{\bar{5}},  \mathbf{10})$ or QCD representations $(r_1,r_2,r_3) =(\mathbf{1}, \mathbf{3},  \mathbf{6})$,  
we have found several charge assignments that lead to $\Delta_{\cancel{PQ}} =12$. These models are perturbative for values of the axion decay constant as low as $4\cdot 10^8\,$GeV and thus give a robust solution to the axion quality problem. For $f_a \sim 10^{12}\,$GeV they can also explain the DM abundance entirely in terms of axions through the misalignment mechanism. Our classification revealed several other models with $n_f=2$ and $n_f=3$ that can also address the quality problem and where the PQ-violating dimension can be higher than 12. For example, we identified GUT models with $n_f=3$ and QCD models with $n_f=2$ or $3$ where $\Delta_{\cancel{PQ}}$ can be as large as 18 and 15 respectively, see Tabs.~\ref{tab:arep} and ~\ref{tab:nf2_rep}. Determining the actual level of protection of these constructions requires a more elaborate analysis that we leave for future work~\cite{Podo:2022gyj}.

Some of the models that we have discussed are close to the edge of the conformal window (for instance those of Section~\ref{sec:GUT} with the choice $N_{\rm DC}=3$), and could feature a walking dynamics. In this case the PQ-violating could have large anomalous dimensions: it would be interesting to understand if such a scenario can improve the level of protection of the PQ symmetry in this class of models or not.

The low-energy dynamics of our models shares all the universal properties of hadronic axion theories. In particular, the GUT models imply an $E/N$ ratio equal to $8/3$, and will be thus tested by on-going and planned experiments aiming at observing QCD axions.
In order to probe the Peccei-Quinn dynamics and distinguish between different proposals, however, further experimental observables are needed.
If $f_a \sim 10^{12} \,{\rm GeV}$ and the breaking of PQ symmetry is induced by $d=12$ local operators generated at $M_{\rm Pl}$, then the models of Section~\ref{sec:nf3} can be tested through the next generation neutron EDM experiments. An experimental indication that the PQ symmetry is accidental could therefore come in the near future, similarly to the case of baryon number and DM symmetry.
Another way to probe the PQ dynamics and distinguish our GUT models from other ones could come from the parametrically lighter states that acquire mass through non-renormalizable interactions. These states may have an impact on cosmological observables and give the tantalizing opportunity of testing the GUT and PQ dynamics through celestial observations.
Besides solving the strong CP problem, the QCD axion could also be the main component of DM, either through the ordinary misalignment mechanism or through a possible realization of the recently proposed kinetic misalignment mechanism~\cite{Co:2019jts,Chang:2019tvx} in the composite scenario.
Furthermore, models with $n_f=2$ predict the existence of an additional massless scalar $s$.
This could be a component of DM, if UV effects generate a mass in the appropriate range and it is populated through a primordial mechanism, and can also give a contribution to dark radiation that, once added to the axion contribution, could be tested by future CMB and large-scale structure observatories.

\section*{Acknowledgements}
We are grateful to Luca Vecchi for useful exchanges on his recent work on composite axions and unified theories~\cite{Vecchi:2021shj}, that appeared while this work was under completion, and for comments on a preliminary version of this article. We would like to thank Lam Hui for useful discussions on cosmological bounds on light scalar dark matter, and Joe Conlon for conversations on axions in String Theory. This research was supported in part by the MIUR under contract 2017FMJFMW (PRIN2017). F.R. acknowledges support from the Dalitz Graduate Scholarship, jointly established by the Oxford University Department of Physics and Wadham College. The work of A.P. is supported by the Simons Foundation Award No.~658906.

\appendix

\section{Discrete symmetries in composite axion models}\label{appendix:discrete}

The global PQ symmetry is broken both explicitly (by the QCD anomaly) and spontaneously (by the $\SUNDC$ condensate). As first pointed out by Sikivie in Ref.~\cite{Sikivie:1982qv}, these symmetry breaking effects often leave unbroken some discrete symmetry groups, which can have important conceptual and phenomenological consequences. In this Appendix we will provide a general analysis of the discrete symmetries related to $\U(1)_{\rm PQ}$ in composite axion models based on a QCD-like confining gauge group.

Let us denote by ${\rm G}_{V}$ the exact global flavour symmetry group preserved by the vacuum. In our models this is a vectorial group of the form ${\rm G}_{V}= \U(1)_{V,1}\times \dots \times \U(1)_{V,n_{irr}}$, where each factor $\U(1)_{V,i}$ acts with charge $Q_{V,i}(\psi_{j})= \delta_{ij}$ and $Q_{V,i}(\chi_{j})= -\delta_{ij}$. As in the main text, we fix the normalization of the $\U(1)_{\rm PQ}$ charges  of the dark fermions $\psi_{i},\chi_{i}$ so that they are coprime integers; with this choice the continuous parameter associated to the symmetry transformation is defined in the range $[0,2\pi)$. In general, the $\U(1)_{\rm PQ}$ PQ group can have a non-vanishing intersection with the unbroken flavor group, $\mathbb{Z}_{V} = \U(1)_{\rm PQ} \cap {\rm G}_{V}$, where $\mathbb{Z}_{V}$ is a discrete group. In other words, the $\SUNDC$ condensates induce the spontaneous breaking
\begin{equation}
\U(1)_{\rm PQ} \longrightarrow \mathbb{Z}_{V}\, .
\end{equation}

The local dark color gauge group $\SUNDC$, which leaves invariant physical states of the Hilbert space, is associated to a global group that instead can act non-trivially on the Hilbert space (see for example~\cite{Strocchi}). Such global group is linearly realized in the confining phase of vector-like theories, and physical states are classified according to its irreducible representations.  One may therefore ask whether this global symmetry plays any role in our analysis of the discrete symmetries related to $\U(1)_{\rm PQ}$.
Since $\U(1)_{\rm PQ}$ commutes with $\SUNDC$, the transformations of such a global $\SUNDC$ that are contained in $\U(1)_{\rm PQ}$ have to be part of the (global) center group $\mathbb{Z}_{N_{\rm DC}}$. The latter acts as a flavour symmetry, and since it is preserved by the vacuum, it must be a subgroup of the vectorial group: $\mathbb{Z}_{N_{\rm DC}} \subset {\rm G}_{V}$. Therefore, there is no need to consider separately transformations of the center $\mathbb{Z}_{N_{\rm DC}}$.

Invariance under the symmetry $\mathbb{Z}_{V}$ restricts the physical domain of the axion. Denoting by $N_{V}$ the order of $\mathbb{Z}_{V}$, one has
\begin{equation}
\dfrac{a}{f_{\rm PQ}} \in \Bigg[0,\dfrac{2\pi}{N_{V}}\Bigg),
\end{equation}
and the domain of PQ shift transformations on the axion is correspondingly restricted.

On the other hand, the anomaly breaks explicitly the $\U(1)_{\rm PQ}$ symmetry down to a discrete group $\mathbb{Z}_{A}$. Since the vectorial group is anomaly free, it follows that $\mathbb{Z}_{A}$ contains $\mathbb{Z}_{V}$ as a subgroup, and in particular its order $N_{A}$ is an integer multiple of $N_{V}$. The quotient group 
\begin{equation}
\mathbb{Z}_{\rm PQ} = \mathbb{Z}_{A}/\mathbb{Z}_{V},
\end{equation}
has order $N_{\rm PQ}=N_{A}/N_{V}$ and is an exact global symmetry group. It is non-linearly realised on the vacuum and acts on the axion as a discrete shift symmetry:
\begin{equation}
\label{eq:ZPQ}
\dfrac{a}{f_{\rm PQ}} \rightarrow \dfrac{a}{f_{\rm PQ}} +\dfrac{k}{N_{\rm PQ}} \dfrac{2\pi}{N_{V}}, \qquad \text{with }k=1,\dots, N_{\rm PQ}\, .
\end{equation}

It is straightforward to show, using the effective chiral lagrangian approach reviewed in~\cite{diCortona:2015ldu}, that (\ref{eq:ZPQ}) is a symmetry of the non-perturbative axion potential. It follows that the axion potential has $N_{\rm PQ}$ distinct minima and the so-called domain wall number $N_{\rm DW}$ equals~$N_{\rm PQ}$.

Let us now determine $N_{A}$ and $N_{V}$ in our models. As in the main text, we use a normalization where ${\rm Tr}(t_{a}t_{b})=\delta_{ab}/2$ for fields in the fundamental representation and use a Dirac notation. Under an anomalous PQ transformation with parameter $\alpha$, the QCD theta angle shifts by\footnote{The $\SU(5)$ GUT case is obtained by replacing the QCD generators with the corresponding GUT generators, obtaining the corresponding $A_{5}$.} $\theta_{0} \rightarrow \theta_{0} -2 A_{3} \alpha$, with $\delta_{ab} A_{3} = N_{\rm DC} {\rm Tr}\left[\{t_{a},t_{b}\} Q_{\rm PQ} \right]$. It follows that $N_{A}= 2 \vert A_{3} \vert$.

In order to compute $N_{V}$, we need to determine $\mathbb{Z}_{V} = \U(1)_{\rm PQ} \cap {\rm G}_{V}$. Notice, first, that fermions $\psi_{i}$ and $\chi_{i}$, with $i=1,\dots, n_{irr}$, have equal PQ charges $Q_{\rm PQ}^{i}$ and opposite $\U(1)_{V,i}$ charges $\pm 1$.
Performing a PQ transformation with parameter $\alpha$ and independent ${\rm G}_{V}$ transformations with parameters $\beta^{i}$, the $\mathbb{Z}_{V}$ is determined by solving the system of equations
\begin{equation}
\begin{split}
&Q_{\rm PQ}^{i} \, \alpha = \beta^{i} + 2 \pi \kappa^{i} \\
&Q_{\rm PQ}^{i} \, \alpha = - \beta^{i} + 2 \pi \kappa'^{\,i},
\end{split}
\end{equation}
with $\kappa^{i}$, $\kappa'^{\,i}$ arbitrary integers. Summing the equations it follows that $ \alpha \, Q_{\rm PQ}^{i} = \pi (\kappa^{i}+\kappa'^{\, i})$, and since the charges $Q_{\rm PQ}^{i}$ are coprime integers, the only non trivial solution in the range $\alpha \in (0,2\pi)$ is $\alpha=\pi$. We find, therefore, that $\mathbb{Z}_{V}= \mathbb{Z}_{2}$ and $N_{V}=2$.

We arrive at our final result
\begin{equation}
N_{\rm DW} = N_{\rm PQ} = \dfrac{N_{A}}{N_{V}}= \vert A_{3} \vert,
\end{equation}
which is always an integer multiple of $N_{\rm DC}$ and, in particular, greater than $1$.

\section{$(\mathbf{1},\mathbf{3},\mathbf{6})$ QCD models}
\label{appendix:QCD3}

In this Appendix we give a brief overview of $n_f=3$ QCD models with $(r_1,r_2,r_3)=(\mathbf{1},\mathbf{\bar{3}},\mathbf{6})$ or $(\mathbf{1},\mathbf{3},\mathbf{6})$, and classify their charge assignments that can solve the quality problem.

The constraints of Sec.~\ref{sec:pt}  imply $3\leq N_{\rm DC} \leq 9$  for these models. The PQ generator takes the form
\begin{equation}
Q_{a}= \text{diag}(+18, + 8\, \mathbb{1}_{3},  - 7\, \mathbb{1}_{6} )\, .
\end{equation}
By counting the residual $\rm{U}(1)$ global symmetries, one can see that there are no additional massless NGBs in the spectrum, so that the low-energy phenomenology is entirely characterized in terms of the axion. The anomaly of $\U(1)_a$ with respect to $\rm{SU}(3)_c$ is given by
\begin{equation}
\langle \partial_{\mu} J_{a}^{\mu} \rangle = - \dfrac{\alpha_{3}}{4\pi} A_{3}\, G \tilde{G}\, ,
\qquad  A_{3} = - 27 N_{\rm DC} \, ,
\end{equation}
so that
\begin{equation}
f_a = \frac{f_{\rm{DC}}}{54 N_{\rm{DC}}}  \qquad \quad N_{DW} =27 N_{DC} \, .
\end{equation}

The anomaly equations (\ref{eq:an3f}) imply the existence of the PQ-violating operator 
\begin{equation}
\mathcal{O} = \psi_1 \chi_1 \big( \psi_2 \chi_2)^2 \psi_3 \chi_3\, ,
\end{equation}
that has dimension 12 and the right flavour structure to generate an axion potential through a single insertion. Solving the strong CP problem for $f_a\sim 10^{12}\,$GeV thus requires the theory to be valid up to the Planck scale. An intermediate scale of Grand Unification in this case would lead to too large corrections to the axion potential, as estimated in Section \ref{sec:pq}, and is therefore not compatible with the axion solution.

High quality models with $\Delta_{\cancel{PQ}}=12$ can be characterized as done in Section~\ref{ssc:exso}. The only difference with respect to the GUT case is the absence of $\rm{SU}(3)_c$ charged scalars, since we only include the SM Higgs field. Therefore, there is no longer a distinction between model dependent and robust solutions to the quality problem. Following the same procedure as in Section~\ref{ssc:exso}, we find the results reported in Table \ref{tab:hqax3}.
\begin{table}
\centering
\begin{tabular}{cl|ccc|ccc|}
&   &  & $N_{\rm{DC}}=3,5,\dots,9 $ &  &  & $N_{\rm{DC}}=4$ \\
& $n_{\rm max}$                    & $10$ & $15$ & $20$  & $10$ & $15$ & $20$    \\
\hline
& AC solutions            & $16$ & $40$ &  $96$ & $16$ & $40$ &  $96$   \\
& HQ axions, $(\mathbf{1},\mathbf{\bar{3}},\mathbf{6})$             & $3$ & $14$ &  $50$  & $3$ & $11$ &  $43$ \\
& HQ axions, $(\mathbf{1},\mathbf{3},\mathbf{6})$             & $4$ & $17$ &  $50$ & $2$ & $15$ &  $43$  \\
& No $d\leq 8$ operators & $0$ & $1$ &  $8$ &$0$ & $0$ &  $1$  \\
\hline
\end{tabular}
\caption{\it Number of $(\mathbf{1},\mathbf{\bar{3}},\mathbf{6})$ and $(\mathbf{1},\mathbf{3},\mathbf{6})$ QCD models with $\Delta_{\cancel{PQ}}=12$ and integer charges satisfying $|p_i|,|q_i| \leq n_{\rm{max}}$. Models are found through a numerical scan of the parameter space for different values of $n_{\rm{max}}$ and $3\leq N_{\rm{DC}}\leq 9$. The counting is made identifying solutions that are equal up to an overall factor or by exchanging $p_i\leftrightarrow q_i$. The first row indicates the total number of independent solutions of the Anomaly Cancellation (AC) conditions of Eq.~(\ref{eq:an3f}), while the following rows list how many of these solutions yield High Quality (HQ) models. The last row shows the number of models with no operators of dimension $d \leq 8$ with a non-trivial flavour structure.}
\label{tab:hqax3}
\vspace{0.1cm}
\end{table}
The three charge assignments with $n_{\rm max} =10$ that solve the quality problem with $(r_1,r_2,r_3)=(\mathbf{1},\mathbf{\bar{3}},\mathbf{6})$ and $3\leq N_{\rm DC} \leq 9$  are: 
\begin{equation}
\begin{aligned}[c]
&(-1,-7,+1;+10,+2,+0)\\
&(+3,-7,+7;+6,+2,-6).\\
\end{aligned}
\qquad \qquad \qquad
\begin{aligned}[c]
&
(-1,+2,+1;+10,-7,+0)\\
& \\
\end{aligned}
\end{equation}
Here and in the following, the $\UoneD$ charges are reported in the format $(p_1,p_2,p_3;q_1,q_2,q_3)$ to keep the notation more compact.
With $(r_1,r_2,r_3)=(\mathbf{1},\mathbf{3},\mathbf{6})$ and $N_{\rm{DC}}=3,5,\dots,9 $, the solutions are
\begin{equation}
 (-1,-7,+1;+10,+2,+0) \,\,\, \quad \quad \quad \quad\quad (-1,+2,+1;+10,-7,+0) 
\end{equation}
and
\begin{equation}
(+3,-7,+7;+6,+2,-6)  \,\,\, \quad \quad \quad \quad \quad  (+4,-4,+0;+5,-1,+1).
\end{equation} 
The first two are also a solution for $(r_1,r_2,r_3)=(\mathbf{1},\mathbf{3},\mathbf{6})$ and $N_{\rm{DC}}=4$.

\section{High quality $(\mathbf{1},\mathbf{\bar{5}},\mathbf{10})$ models}
\label{appendix:hq}

Following the analysis of Section~\ref{ssc:exso}, we present the charge assignments for $n_f=3$ GUT models with $(r_1,r_2,r_3)=(\mathbf{1},\mathbf{\bar{5}},\mathbf{10})$ or $(\mathbf{1},\mathbf{5},\mathbf{10})$ and $\Delta_{\cancel{PQ}}=12$.
We report all the high-quality solutions with $n_{\rm{max}} = 10$ classified in Table \ref{tab:hqax}.

For $(r_1,r_2,r_3)=(\mathbf{1},\mathbf{\bar{5}},\mathbf{10})$ and $5\leq N_{\rm{DC}}\leq 11$, the robust solutions are those of Eq.~(\ref{eq:solex}), while there are none for $N_{\rm{DC}} = 4$. For $4\leq N_{\rm{DC}} \leq 11$, there are the following high-quality solutions with a minimal GUT scalar sector:\footnote{For $5\leq N_{\rm{DC}} \leq 11$,  the sixth solution is also robust, i.e. it belongs to the list in Eq.~(\ref{eq:solex}).} 
\begin{equation}
\begin{aligned}[c]
&(-5,+6,+8;+10,-9,-7)\\
&(-3,+4,+6;+8,-7,-5) \\
&(-1,+2,+4;+6,-5,-3) \\
&(+5,-2,+4;+10,-7,-1) \\
&(+7,-4,+2;+8,-5,+1).\\
\end{aligned}
\qquad \qquad \qquad
\begin{aligned}[c]
&(-4,+5,+7;+9,-8,-6)\\
&(-2,+3,+5;+7,-6,-4)\\
&(+0,-7,+7;+5,+4,-6)\\
&(+5,+1,-5;+10,-10,+8)\\
& \\
\end{aligned}
\end{equation}
Finally, for $5\leq N_{\rm{DC}} \leq 11$ and with a minimal GUT sector we found the additional high-quality solutions 
\begin{equation}
\begin{aligned}[c]
&(-5,+3,+2;+10,-6,-1)\\
&  (-4,+5,-6;+9,-8,+7) \\
&  (-2,+3,-4;+7,-6,+5) \\
&  (+5,-6,+2;+10,-3,+1) \\
&(+5,-2,-1;+10,-7,+4) \\
\end{aligned}
\qquad \qquad \qquad
\begin{aligned}[c]
&  (-5,+6,-7;+10,-9,8)\\
&  (-3,+4,-5;+8,-7,+6)\\
&  (-1,+2,-3;+6,-5,+4) \\
&  (+5,-3,+2;+10,-6,+1)\\
&  (+7,-4,+1;+8,-5,+2). \\
\end{aligned}
\end{equation}

With $(r_1,r_2,r_3)=(\mathbf{1},\mathbf{5},\mathbf{10})$, there are fewer solutions compared to the other choice of representations. With $5\leq N_{\rm{DC}} \leq 11$, the robust ones are the last three of Eq.~(\ref{eq:solex}). Assuming a minimal GUT sector one has the additional high-quality solutions
\begin{equation}
\begin{aligned}[c]
&  (-5,+3,-1;10,-6,+2)\\
&  (-4,+5,-6;+9,-8,+7) \\
&  (-2,+3,-4;+7,-6,+5) \\
&  (+0,+1,-2;+5,-4,+3) \\
&  (+5,-3,+2;+10,-6,+1) \\
&  (+7,-4,+1;+8,-5,+2). \\
\end{aligned}
\qquad \qquad \qquad
\begin{aligned}[c]
&  (-5,+6,-7;+10,-9,+8)\\
&  (-3,+4,-5;+8,-7,+6)\\
&  (-1,+2,-3;+6,-5,+4) \\
&  (+5,-6,+2;+10,-3,+1)\\
&  (+5,-2,-1;+10,-7,+4) \\
& \\
\end{aligned}
\end{equation}
With $N_{\rm{DC}} = 4$, there are no robust solutions, and only two high-quality ones with a minimal GUT sector:
\begin{equation}
\begin{aligned}[c]
& (+2,+3,-5;+3,-6,+6),\\
\end{aligned}
\qquad \qquad \qquad
\begin{aligned}[c]
&(+0,+4,-6;+5,-7,+7).\\
\end{aligned}
\end{equation}

\section{Randall's model}\label{appendix:Randall}

The model by Randall~\cite{Randall:1992ut} is based on a product gauge group $\mathcal{G} = \SUNDC \times \GW \times \GSM$, with a non-abelian weak factor $\GW = \SUm$, and is defined by the chiral field content of Table~\ref{tab:randall}.
\begin{table}[t]
\centering
\begin{tabular}{lccc|c}
& $\SUNDC$ & $\SU(m)$ & $\rm \GSM$ & $\U(1)_{\rm PQ}$ \\
\cline{1-5} 
\rule{0pt}{2.4ex}$\psi_1$ & $\square$ &  $\mathbf{m}$ & $r$ & $+1$\\
$\psi_{2,i}$ & $\square$ &  $\mathbf{\overline{m}}$ & $\mathbf{1}$ & $-1$\\         
\cline{1-5}        
\rule{0pt}{2.4ex}$\chi_{1,j}$ & $\bar{\square}$ &  $\mathbf{1}$  & $\bar{r}$ & $+1$\\        
$\chi_{2,k}$ & $\bar{\square}$ &  $\mathbf{1}$  & $\mathbf{1}$ & $-1$\\ 
\end{tabular}
\caption{\it Composite axion model of Ref.~\cite{Randall:1992ut}. In the original work only QCD interactions were considered, with $\GSM=\SU(3)_{c}$ and $r= \mathbf{3}$; the minimal GUT extension is obtained by taking $\GSM=\SU(5)$ and $r=\mathbf{5}$. The fermions $\psi_{2},\chi_{1},\chi_{2}$ appear in multiple copies, labelled by the indices $i=1,\dots,\dim(r)$, $j=1,\dots,m$, $k=1,\dots,m\cdot\dim(r)$.}
\label{tab:randall}
\end{table}
The original work of Ref.~\cite{Randall:1992ut} and the analysis of Dobrescu of Ref.~\cite{Dobrescu:1996jp} considered the pure QCD case where $\GSM=\SU(3)_{c}$, but the model can be straightforwardly generalized to a GUT scenario by replacing the fundamental representation of $\SU(3)_{c}$ with that of $\SU(5)$.

The multiplicity of dark flavours, as defined in Sec.~\ref{sec:mb}, is $N_{f}=6 m \;(10m)$ for the QCD (GUT) case, and $\Delta N_{f}^{(\rm QCD)}= m N_{\rm DC}$, where $m$ is the number of colors of the non-abelian $\SU(m)$ factor. Compared to the models of Section~\ref{sec:nf2} we see that the effective number of QCD flavours is the same for $m=2$, however $N_{f}$ is always larger in the model by Randall. The requirements of confinement~\eqref{eq:dc_coupling_conf} and perturbativity~\eqref{eq:qcd_coupling} are compatible with $m\leq 4 (3)$ for the QCD (GUT) case if $f_a= 10^{9}\, \rm GeV$, and with $m\leq 5 (4)$ for the QCD (GUT) case if $f_a= 10^{12}\, \rm GeV$. These bounds have been obtained by estimating $\Lambda_{\rm DC} = 4\pi f_{\rm DC}/\sqrt{N_{\rm DC}}$ and using $A_3 = N_{\rm DC} m$ to write $f_{\rm DC}=2 N_{\rm DC}m f_a$.

As already noticed in Ref.~\cite{Dobrescu:1996jp}, the leading PQ-violating operators that generate a potential for the axion at the level of a single insertion are of the form $(\psi_{1}\chi_{1,j})^{m-n}(\bar{\psi}_{2,i}\bar{\chi}_{2,k})^{n}$, with $n=0,\dots,m$ ($n\neq m/2$), and have dimension $3m$. According to our estimates of Sec.~\ref{sec:pq}, in the QCD case the confinement and perturbativity constraints are compatible with a solution of the quality problem if there are no intermediate new scales below the Planck scale, both for $f_{a} = 10^{9}\, \rm GeV$ and for $f_{a} = 10^{12}\, \rm GeV$. The more sophisticated analysis of perturbativity performed in Ref.~\cite{Dobrescu:1996jp} suggests that the axion solution is actually spoiled for values of the Peccei-Quinn scale $f_{\rm DC} \gtrsim 10^{11}\,$GeV.

One can wonder whether multiple insertions can induce PQ-violating effects that are more dangerous than those from single insertions. We find that multiple insertions are less dangerous than single ones in the case of QCD models. On the other hand, in the GUT case the leading PQ-violating operators have dimension 7 and include a GUT scalar $\Phi$ in the fundamental representations of $\SU(5)$; they are of the form $\mathcal{O}_{1}= \psi_{1}\psi_{2}(\chi_{1})^{2} \Phi$ and $ \mathcal{O}_{2}= \psi_{1}\psi_{2}(\chi_{2})^{2} \Phi^{\dagger}$. According to the criteria discussed in Section~\ref{sec:spurion}, these operators are harmless at the level of a single insertion but give rise to a potential for the axion when combined in a double insertion. On dimensional ground this effect is equivalent to the insertion of a single operator of effective dimension $10$, and thus provides the leading PQ-breaking effect when $m\geq4$. This suggests that GUT models can solve the quality problem only for low values of the axion decay constant, $f_a \sim 10^9\,$GeV.

\end{document}